\documentclass[pre,onecolumn,showpacs,preprintnumbers,pra,showkeys,amssymb]{revtex4}
\usepackage{amsfonts}
\usepackage{amsmath}
\usepackage{graphicx}
\usepackage{dcolumn}
\usepackage{epsfig}
\usepackage{color}
\usepackage{hyperref}
\usepackage{hyphenat}
\usepackage{bm}
\usepackage{float}
\newcommand{\be}{\begin{equation}}
\newcommand{\beast}{\begin{equation*}}
\newcommand{\ee}{\end{equation}}
\newcommand{\eeast}{\end{equation*}}
\newcommand{\br}{\begin{eqnarray}}
\newcommand{\brast}{\begin{eqnarray*}}
\newcommand{\er}{\end{eqnarray}}
\newcommand{\erast}{\end{eqnarray*}}
\newcommand{\bse}{\begin{subequations}}
\newcommand{\ese}{\end{subequations}}

\newcommand{\bd}{\begin{displaymath}}
\newcommand{\ed}{\end{displaymath}}

\newcommand{\bfig}{\begin{figure}}
\newcommand{\efig}{\end{figure}}

\begin{document}
\title{Modeling a vehicular traffic network. Part I}
\author{D. Otero$^{1}$, D. Galetti$^{2}$, and S. S. Mizrahi$^{3}$}
\email{dinootero@fibertel.com.ar, diogaletti@hotmail.com,
salomon@df.ufscar.br}
\affiliation{$^{1}$Facultad Regional General Pacheco, UTN, Buenos Aires, Argentina \\
$^{2}$Instituto de F\'{\i}sica Te\'{o}rica, UNESP, S\~{a}o Paulo, SP, Brasil 
\\
$^{3}$Departamento de F\'{\i}sica, CCET, Universidade Federal de S\~ao
Carlos, 13565-905, S\~ao Carlos, SP, Brazil}
\date{\today}

\begin{abstract}
We propose three models for the traffic of vehicles within a network formed by 
sites (cities, car-rental agencies, parking lots, etc.) and connected by two-way 
arteries (roads, highways), that allow forecasting the vehicular flux in a 
sequence of $n$ consecutive steps, or units of time. An essential approach consists 
in using, as an \emph{a priori} information, previous observations and measurements. 
The formal tools used in our analysis consists in: (1) associating a digraph to the 
network where the edges correspond to arteries and the vertices with loops 
represent the sites. (2) From an initial set of numbers, that are the distribution   
of vehicles within the network, we construct a matrix that we transform into a 
stochastic matrix (SM) by normalizing the rows, whose entries are now  
transition probabilities. This matrix becomes the generator of the evolution of 
the traffic flow. And (3), we use the Perron-Frobenius theory for a formal analysis. 
We investigate three models: (a) a closed four-site network having a conserved 
number of vehicles; (b) to this network we add an influx and an outflux of vehicles 
to characterize an open system; asymptotically, $n \rightarrow \infty$, the SM 
raised to the power $n$ goes to a unique stationary matrix. And (c), we construct a 
nonlinear model because the formal structure permits the existence of several 
($L$) stationary states for the distribution of vehicles at each site, that alternate 
cyclically with time. Each state represents the traffic for $L$ different moments. 
These models were used to analyze the traffic in a sector of the city of Tigre, 
located in the province of Buenos Aires, Argentina. The results are presented in a 
following paper.
\end{abstract}

\maketitle




%
%
\section{Introduction} 
%
Understanding the structure and systematic of the flow of vehicles along 
a network of arteries is a crucial point for planning the construction of 
connections (roads and highways) linking sites, as is the need to forecast 
the vehicular flow for traffic control. For that aim the observation 
and the gathering of data are essential practices that have to go along 
with theoretical approaches. Mathematical modeling and numerical simulations 
are essential procedures to grasp at the traffic flow and jam problems 
in order to recommend solutions for the urban and inter-city vehicular mobility. 
Methods have been developed since the 1940 decade, see for instance the seminal 
papers \cite{pipes1953,lighthill1955} and references therein. Further studies 
were done in several places uninterruptedly and many articles and reports 
were published on the subject, among which we cite, for instance, the 
Refs. \cite{bando1994,bando19951,bando19952,kerner1997,helbing1998,nagatani1998,
deangelis1999,aw2000,helbing2001,nagatani2002,bellomo2002,
greenberg2004,kisselev2004,gasser2004,orosz2005,li2006,siebel2006,
guanghan2009,bressan2011,bressan2012,goh2012,forstall2013,
bressan2014,canec2015,bressan2016,piccoli,morein}, and the books 
\cite{kerner2004,manhke2009}. 

The success of a model depends essentially on its capability to describe the 
present traffic dynamics as well as to forecast its trend. There are approaches 
that make use of fluid dynamics and differential equations 
\cite{bando19951,helbing1998} whereas others utilize the $n$-step evolution 
of a stochastic matrix   
\cite{manhke2009,ching2006,woess-PF}. Here we consider the vehicular circulation 
occurring essentially inter-sites, and by sites we mean towns, parking lots, 
car-rental agencies, etc. 

In this study we do not describe the features of the vehicles, as mean size, 
mean speed, stopping distance nor the characteristics of the arteries as length, 
width and number of lanes. It consists in schematizing a traffic network imaged 
by a digraph, as shown in Fig. \ref{fig01}, that could be scaled up to higher 
complexity. The mathematical elements we utilize are essentially the digraph 
methods and the Perron-Frobenius theory for stochastic matrices, where each 
$n$-step entry is a Markov chain. We first present two linear models: 
(1) a network consisting of four sites, each one connected to all the 
others by two-way arteries. The total number of circulating vehicles is 
assumed to be constant in time. (2) Thereafter, we extend that network 
introducing an input of new vehicles and an output of old ones; 
see the digraph in Fig. \ref{fig03}. To each digraph one associates a 
square matrix $\mathbb{A}$ whose entries contain the information obtained 
empirically, namely, the number of vehicles established by counts. 
To forecast the distribution of the vehicles at future moments in the 
network we normalize the rows of $\mathbb{A}$ to 1, thus getting a matrix  
$\mathbb{M}$. Its entries are the fractions of vehicles in sites and 
roads. In this way, the predictions will be probabilistic and we assume 
that the daily change of the distribution obeys an evolution law based 
on an $n$-step process -- $n$ standing for the discretized time. This 
approach constitutes an adaptation of a model proposed in Ref. 
\cite{harary1966} for human migrations. We present the theory, work out 
illustrative numerical examples and analyze the results. Additionally, 
we extend our analysis by examining a third model, (3) a nonlinear 
$n$-step process; by non-linearity we mean that some entries of matrix 
$\mathbb{M}$ depend on $n$. With a specific choice of periodic functions 
for the entries we verify the existence of several stationary r\'{e}gimes 
(instead of a single one as it occurs in the linear models), such that 
the vehicle distributions change cyclically instead of remaining constant. 
This nonlinear approach enables more detailed predictions about the traffic 
dynamics because it permits to slice, for instance, a 24-hour day into 
several sub-periods of traffic observation in the stationary r\'{e}gimes. 

The present theoretical study was used to analyze the urban traffic 
in a selected sector of Tigre, a city located in the province of 
Buenos Aires, Argentina. The raw data are counts of circulating vehicles 
as recorded by cameras positioned in several intersections. The methodology, 
analysis and comparison between raw data and the utilized model are presented 
in a following paper, to be referred as Part II. 
%
\section{Model I: Inter-site circulation of vehicles between 
without input or output.}
\begin{figure}[H]
\centering
\includegraphics[height=2.4in, width=3.0in]{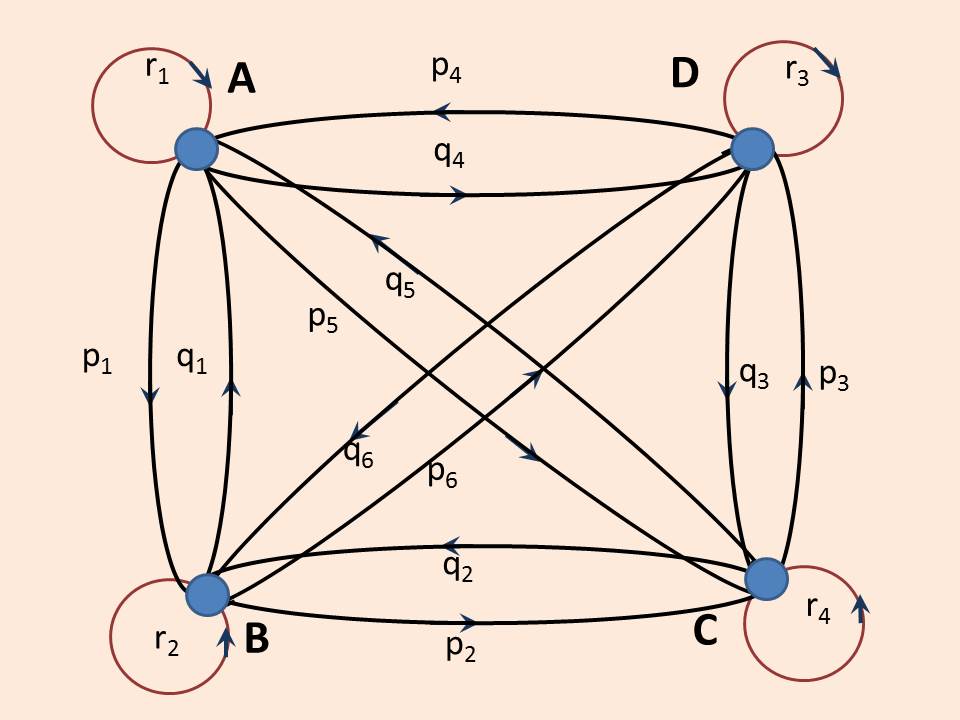}
\caption{\small {The network has sites (cities, parking lots, intersections, 
etc.), represented by the vertices with respective loops, and edges representing 
the arteries that connect the vertices. The $r_{i}$'s, at the loops, represent 
the fractions of the vehicles present in the sites whereas $p_{i}$'s and 
$q_{i}$'s, at the edges, stand for the fractions of vehicles traveling from one 
artery to the other, depending on the directions of the arrows.}}
\label{fig01}
\end{figure}

%
The first model consists of four sites, to which we attribute the 
letters $A, B, C, D$, connected by arteries (roads, highways, etc.) and it  
contains a fixed number of vehicles; in short it is a conservative network. 
Pictorially we represent the network by a digraph, as shown in Fig. \ref{fig01}, 

\noindent where each \emph{loop}, associated with a \emph{vertex}, corresponds 
to a site that accommodates a certain number of vehicles, whereas each 
\emph{edge} is a link between the vertices that contains a number of vehicles 
in transit from one site to the other along the direction of the arrows. 
To the digraph of Fig. \ref{fig01} one associates the array shown in Table 
\ref{A0}, 
\begin{table}[htbp]
\begin{tabular}{|c||llll||l|}
\hline
-- & $A$ & $B$ & $C$ & $D$ & Sum of the lines entries \\ \hline\hline
$A$ & $R_{1}$ & $P_{1}$ & $P_{5}$ & $Q_{4}$ & \multicolumn{1}{c|}{${\ U}_{1}$%
} \\ \hline
$B$ & $Q_{1}$ & $R_{2}$ & $P_{2}$ & $P_{6}$ & \multicolumn{1}{c|}{${\ U}_{2}$%
} \\ \hline
$C$ & $Q_{5}$ & $Q_{2}$ & $R_{3}$ & $P_{3}$ & \multicolumn{1}{c|}{${\ U}_{3}$%
} \\ \hline
$D$ & $P_{4}$ & $Q_{6}$ & $Q_{3}$ & $R_{4}$ & \multicolumn{1}{c|}{${\ U}_{4}$%
} \\ \hline\hline
Sum of the columns entries & ${\ T}_{1}$ & ${\ T}_{2}$ & ${\ T}_{3}$ & ${\ T}%
_{4}$ & \multicolumn{1}{c|}{--} \\ \hline\hline
-- & \multicolumn{4}{c||}{$Y=\sum_{i=1}^{4}T_{i}$} & \multicolumn{1}{c|}{$%
Y=\sum_{i=1}^{4}U_{i}$} \\ \hline
\end{tabular}%
\caption{{\small {The labels $A$, $B$, $C$ and $D$ specify the
vertices of the digraph, the entries $R_{i}$ are the number of vehicles in 
the sites, and the $Q_{i}$'s and $P_{i}$'s are the numbers of vehicles on 
the roads, running from one vertex to another. The sum of the entries of 
each row and column are $U_{i}$, and $T_{i}$ respectively. $Y$ is the total 
number of vehicles in the network.}}}
\label{A0}
\end{table}
the sum of the entries of each row, $U_{i}$, and the sum of the entries of
each column, $T_{i}$, are presented in Table \ref{A0}, and in the last row
one finds the sums for the total number of vehicles $Y$ in the network.

The core of the array of Table \ref{A0} is a $4\times 4$ matrix 
\begin{equation}
\mathbb{A}=\left( 
\begin{array}{cccc}
R_{1} & P_{1} & P_{5} & Q_{4} \\ 
Q_{1} & R_{2} & P_{2} & P_{6} \\ 
Q_{5} & Q_{2} & R_{3} & P_{3} \\ 
P_{4} & Q_{6} & Q_{3} & R_{4}%
\end{array}%
\right) ,  \label{A1}
\end{equation}
whose entries are the number vehicles observed and counted in an \emph{ad hoc} 
interval of time or an average over several previous observations. 

Assuming that we want to forecast the evolution of the distribution of vehicles
in the network at a daily basis, for instance, we adopt the causal 
interpretation based on the hypothesis that the vehicles distribution evolves 
according to a $n$-step process: (\emph{a}) The entries in the main diagonal in
matrix (\ref{A1}), $A_{ii}=R_{i}$, stand for the number of vehicles at site $i$; %
(\emph{b}) the off diagonal entries $A_{ij}$ ($i\neq j$) are for the number
of vehicles that left site $i$ and are on their way to site $j$. For a period of
24 hours the sum of the entries of row $i$, $U_{i}=\sum_{j}A_{ij}$ represents 
the number of vehicles that remained in or left the site $i$ for the other 
sites. Complementarily, $T_{j}=\sum_{i}A_{ij}$ represents the number of vehicles 
that are in site $j$ plus those that have departed from the other sites and are on 
their way to arrive at it. So, we may consider that matrix (\ref{A1}) represents 
a continuous observation for a 24-hour period, or an average over sub-periods of 
observations, for instance, six minutes sample at every two hours. 
%
\section{The Stochastic Matrix }
%
By normalizing the rows of matrix (\ref{A1}) we construct a stochastic matrix 
(SM) to be adopted as the generator of the evolution that could describe and 
forecast the number of vehicles in the sites plus those in transit. Writing 
the parameters for the first row as
\begin{equation}
r_{1}=\frac{R_{1}}{U_{1}},\quad p_{1}=\frac{P_{1}}{U_{1}},\quad p_{5}=\frac{%
P_{5}}{U_{1}},\quad q_{4}=\frac{Q_{4}}{U_{1}},  \label{normline}
\end{equation}
the same goes for the other rows, which are used to construct the 
transition probability matrix  
\begin{equation}
\mathbb{M}=\left( 
\begin{array}{cccc}
r_{1} & p_{1} & p_{5} & q_{4} \\ 
q_{1} & r_{2} & p_{2} & p_{6} \\ 
q_{5} & q_{2} & r_{3} & p_{3} \\ 
p_{4} & q_{6} & q_{3} & r_{4}%
\end{array}%
\right)\ .  \label{A2}
\end{equation}
The sum of the (non-negative) entries of each row is $1$ whereas the sum of
the entries of each column is not necessarily $1$, however, if additionally
the sum of the entries of each column happens also to be $1$ then the matrix
is said to be \emph{doubly stochastic}. In more realistic instances the
entries of a SM, as (\ref{A2}), can be constructed using empirical data 
collected from previous observations and 
\begin{equation}
\mathbb{U}^{\mathrm{T}}\left( 0\right) =\left( 
\begin{array}{cccc}
U_{1}\left( 0\right) & U_{2}\left( 0\right) & U_{3}\left( 0\right) & 
U_{4}\left( 0\right)%
\end{array}%
\right) ,  \label{A22}
\end{equation}
represents an initial state for the distribution of vehicles; the superscript 
$\mathrm{T}$ stands for transposition. The $n$-step process evolves of state 
(\ref{A22}) as
\begin{equation}
\mathbb{U}^{\mathrm{T}}(n)= \mathbb{U}^{\mathrm{T}}(0)\mathbb{M}^{n} \ ,
\label{A24}
\end{equation}
that is presumed to forecast the distribution of vehicles at the $n$-th moment.   
%
\subsection{Properties of stochastic matrices (SM)}
%
\textbf{(1)} The sum of the components of the vector $\mathbb{U}(n)$ 
is a conserved quantity, 
\begin{equation}
\sum_{i}U_{i} {\left( n\right) }=\sum_{i}U_{i}{\left( 0\right) } \quad
n=1,2,3,...\ .  \label{A25}
\end{equation}
because $\mathbb{M}^{n}$ is a SM \cite{harary1966}.
\vspace{2mm}

\textbf{(2)} The SM $\mathbb{M}\geq 0$, of dimension $N \times N$, 
has eigenvalues $\lambda _{1}=1$ and $\left\vert \lambda
_{2}\right\vert ,\left\vert \lambda _{3}\right\vert ,...,\left\vert \lambda
_{N}\right\vert <1$, therefore $\lim_{n\rightarrow \infty }\left\vert
\lambda _{k}\right\vert ^{n}=0$, for $k\neq 1$. Eigenvalue $\lambda _{1}$ is 
known in the literature as the Perron-Frobenius (PF) eigenvalue \cite{ching2006}. 
\vspace{2mm}

\textbf{(3)} If all the eigenvalues of $\mathbb{M}$ have multiplicity 
1 with linearly independent eigenvectors $\mathbb{X}%
^{\left( 1\right) }$, $\mathbb{X}^{\left( 2\right) }$, ..., $\mathbb{X}%
^{\left( N\right) }$, we construct the matrix 
\begin{equation}
\mathbb{Q}=\left( 
\begin{array}{cccc}
\mathbb{X}^{\left( 1\right) } & \mathbb{X}^{\left( 2\right) } & \cdots & 
\mathbb{X}^{\left( N\right) }%
\end{array}%
\right) ,  \label{A3}
\end{equation}
and write the decomposition of $\mathbb{M}$ and its powers as 
\begin{equation}
\mathbb{M}=\mathbb{QDQ}^{-1},\quad \text{and\quad }\mathbb{M}^{n}=\mathbb{QD}%
^{n}\mathbb{Q}^{-1},  \label{A8}
\end{equation}
where $\mathbb{D}$ is a diagonal matrix whose entries are the eigenvalues
and $\mathrm{Tr}\mathbb{M}^n = \sum_{k=1}^{N}\left(\lambda _{k}\right)^n$ 
since the eigenvalue equation $%
\mathbb{M}^{n}\mathbb{X}^{\left( k\right) }=\left( \lambda _{k}\right) ^{n}%
\mathbb{X}^{\left( k\right) }$ holds. The spectral decomposition of $\mathbb{M}%
^{n}$ in terms of the stationary state and the decaying modes is 
\begin{equation}
\mathbb{M}^{n}=\mathbb{C}_{1}+\sum_{l=2}^{N}\left( \lambda _{l}\right) ^{n}%
\mathbb{C}_{l}.  \label{B28}
\end{equation}
The matrix $\mathbb{C}_{1}$ (associated with the PF eigenvalue) is a stationary 
SM; the other matrices $\mathbb{C}_{l}$, $l>1$, are not stochastic, having however 
the property $\mathrm{Tr}\left( \mathbb{C}_{l}\right) =1$. Since $\left\vert \lambda
_{l}\right\vert <1$, then $\lim_{n\rightarrow \infty }\mathbb{M}^{n}=\mathbb{%
C}_{1}$, which is asymptotically irreversible, $\mathrm{%
\det }\left( \mathbb{C}_{1}\right) =0$. The eigenvalues $\lambda _{l} $ can
be associated with characteristic relaxation times: we write $\lambda _{l}=%
\mathrm{sgn}\left( \lambda _{l}\right) $ $\left\vert \lambda _{l}\right\vert 
$, where $\mathrm{sgn}\left( \cdot \right) $ is the sign of the argument.
Therefore, we define the characteristic decay time as  
$T_{l}=-\left( \ln \left\vert \lambda _{l}\right\vert \right) ^{-1}$, such
that Eq. (\ref{B28}) can be written as 
\begin{equation}
\mathbb{M}^{n}=\mathbb{C}_{1}+\sum_{l=2}^{N}\left[ \mathrm{sgn}\left(
\lambda _{l}\right) \right] ^{n}\mathbb{C}_{l}\exp \left( -\frac{n}{T_{l}}%
\right) .  \label{B29}
\end{equation}
Thereafter, one can write the evolution of the vector $\mathbb{U}^{\mathrm{T}%
}\left( 0\right) $, $\mathbb{U}^{\mathrm{T}}\left( n\right) =\mathbb{U}^{%
\mathrm{T}}\left( 0\right) \mathbb{M}^{n}$ as 
\begin{equation}
\mathbb{U}^{\mathrm{T}}\left( n\right) =\mathbb{U}^{\mathrm{T}}\left(
0\right) \mathbb{C}_{1}+\sum_{l=2}^{N}\left[ \mathrm{sgn}\left( \lambda
_{l}\right) \right] ^{n}\left( \mathbb{U^{\mathrm{T}}}\left( 0\right) 
\mathbb{C}_{l}\right) \exp \left( -\frac{n}{T_{l}}\right) ,  \label{B30}
\end{equation}
where $\mathbb{U}^{\mathrm{T}}\left( 0\right) \mathbb{C}_{1}=\lim_{n%
\rightarrow \infty }\mathbb{U}^{\mathrm{T}}\left( n\right) $ is the
asymptotic distribution of vehicles and $\left[ \mathrm{sgn}\left( \lambda _{l}\right) %
\right] ^{n}\times \left( \mathbb{U^{\mathrm{T}}}\left( 0\right) \mathbb{C}%
_{l}\right) \exp \left( -n/T_{l}\right) $ are ($N-1$) partial distributions
in the transient regimes and the matrices can have negative entries,
although the entries of the vectors $\mathbb{U}^{\mathrm{T}}\left( 0\right) 
\mathbb{C}_{1} $ and $\mathbb{U}^{\mathrm{T}}\left( n\right) $ are positive.
We illustrate the theory through an example. \vspace{2mm}
%

\noindent \underline{\textbf{Example 1:}} Let us consider the SM 
%
\begin{equation}
\mathbb{M}=\left( 
\begin{array}{ccc}
 1/2 & {0} &  1/2 \\ 
 1/4 &  1/2&  1/4\\ 
 4/10&  3/10&  3/10%
\end{array}%
\right) \ ,  \label{C1}
\end{equation}
whose eigenvectors and eigenvalues are 
\begin{subequations}
\label{C2}
\begin{eqnarray}
\mathbb{X}^{\left( 1\right) } &=&\left( 
\begin{array}{r}
{0.577350} \\ 
{0.577350} \\ 
{0.577350}%
\end{array}%
\right) ,\quad {\ \lambda }_{1}{\ =1},\quad \text{{\ stationary mode,}}
\label{C1a} \\
\mathbb{X}^{\left( 2\right) } &=&\left( 
\begin{array}{r}
{0.576331} \\ 
-{0.802915} \\ 
-{0.152215}%
\end{array}%
\right) ,\quad {\ \lambda }_{2}{\ =0.367945},{\ \quad }\text{{\ decaying
mode,}}  \label{C2c} \\
\mathbb{X}^{\left( 3\right) } &=&\left( 
\begin{array}{r}
{0.660268} \\ 
{0.039495} \\ 
-{0.749991}%
\end{array}%
\right) ,\quad {\ \lambda }_{3}{\ =-0.067945,\quad }\text{{\ decaying mode.}}
\label{C2e}
\end{eqnarray}
\end{subequations}
The matrix $\mathbb{M}$ is regular because $\mathbb{M}^{2}$ is already
positive (all entries are positive).  We construct the matrix $\left( \begin{array}{ccc}
\mathbb{X}^{\left( 1\right) } & \mathbb{X}^{\left( 2\right) } & \mathbb{X}%
^{\left( 3\right) } \end{array}\right)$,
\begin{equation}
\mathbb{Q}=\left( 
\begin{array}{rrr}
{0.577350} & {0.576331} & {0.660268} \\ 
{0.577350} & -{0.802915} & {0.039495} \\ 
{0.577350} & -{0.152215} & -{0.749991}%
\end{array}%
\right) ,  \label{C4}
\end{equation}
and $\mathbb{D}=\mathrm{Diag}[1.0,0.367945,-0.067945]$. 
The matrix $\mathbb{M}^{n}=\mathbb{QD}^{n}\mathbb{Q}^{-1}=\left( 
\begin{array}{ccc}
\mathbb{Y}^{\left( 1,n\right) } & \mathbb{Y}^{\left( 2,n\right) } & \mathbb{Y%
}^{\left( 3,n\right) }%
\end{array}%
\right)$ so that the three columns are 
\begin{equation}
\mathbb{Y}^{\left( 1,n\right) }=\underbrace{\frac{{\ 11}}{{\ 27}}\left( 
\begin{array}{c}
{\ 1} \\ 
{\ 1} \\ 
{\ 1}%
\end{array}%
\right) }_{\text{stationary mode}}+\underbrace{\left( 
\begin{array}{r}
{0.304793} \\ 
{-0.424622} \\ 
{-0.326908}%
\end{array}%
\right) \left( {\lambda _{2}}\right) ^{n}+\left( 
\begin{array}{r}
{0.287800} \\ 
{0.017215} \\ 
{-0.326908}%
\end{array}%
\right) \left( {\ \lambda _{3}}\right) ^{n}}_{\text{ decaying modes}}
\label{C11}
\end{equation}
\begin{equation}
\mathbb{Y}^{\left( 2,n\right) }=\underbrace{\frac{{\ 2}}{{\ 9}}\left( 
\begin{array}{c}
1 \\ 
1 \\ 
1%
\end{array}%
\right) }_{\text{stationary mode}}+\underbrace{\left( 
\begin{array}{r}
{-0.544452} \\ 
{0.758503} \\ 
{0.143795}%
\end{array}%
\right) \left( {\ \lambda _{2}}\right) ^{n}+\left( 
\begin{array}{r}
{0.322230} \\ 
{0.019275} \\ 
{-0.366017}%
\end{array}%
\right) \left( {\ \lambda _{3}}\right) ^{n}}_{\text{decaying modes}}
\label{C12}
\end{equation}
\begin{equation}
\mathbb{Y}^{\left( 3,n\right) }=\underbrace{\frac{{\ 10}}{{\ 27}}\left( 
\begin{array}{c}
1 \\ 
1 \\ 
1%
\end{array}%
\right) }_{\text{stationary mode}}+\underbrace{\left( 
\begin{array}{r}
{0.239659} \\ 
{-0.333880} \\ 
{-0.063296}%
\end{array}%
\right) \left( {\ \lambda _{2}}\right) ^{n}+\left( 
\begin{array}{r}
{-0.610029} \\ 
{-0.036490} \\ 
{0.692926}%
\end{array}%
\right) \left( {\ \lambda _{3}}\right) ^{n}}_{\text{decaying modes}}\ .
\label{C13}
\end{equation}
We further write the spectral decomposition of matrix (\ref{C1}) in terms of
the modes or regimes $\mathbb{M}^{n}\mathbb{=C}_{1}+\mathbb{C}_{2}\left( {\
\lambda _{2}}\right) ^{n}+\mathbb{C}_{3}\left( {\ \lambda _{3}}\right) ^{n}$, 
where the three matrices and their determinants are 
\begin{equation}
\mathbb{C}_{1}=\left( 
\begin{array}{ccc}
11/27 & 2/9 & 10/27 \\ 
11/27 & 2/9 & 10/27 \\ 
11/27 & 2/9 & 10/27%
\end{array}%
\right) ;\quad \det \mathbb{C}_{1}=0,  \label{C15}
\end{equation}
\begin{equation}
\mathbb{C}_{2}=\left( 
\begin{array}{rrr}
{\ 0.304793} & {\ -0.544452} & {\ 0.239659} \\ 
{\ -0.424622} & {\ 0.758503} & {\ -0.333880} \\ 
{\ -0.326908} & {\ 0.143795} & {\ -0.063296}%
\end{array}%
\right) ;\quad \det \mathbb{C}_{2}=1.\,\allowbreak 076\,11\,\times 10^{-7}%
\text{,}  \label{C16}
\end{equation}
\begin{equation}
\mathbb{C}_{3}=\left( 
\begin{array}{rrr}
{\ 0.287800} & {\ 0.322230} & {\ -0.610029} \\ 
{\ 0.017215} & {\ 0.019275} & {\ -0.036490} \\ 
{\ -0.326908} & {\ -0.366017} & {\ 0.692926}%
\end{array}%
\right) ;\quad \det \mathbb{C}_{3}=3.\,\allowbreak 245\,95\times 10^{-13}.
\label{C17}
\end{equation}
While matrix $\mathbb{C}_{1}$ does not have an inverse, $\mathbb{C}_{2}$ and 
$\mathbb{C}_{3}$ do have, but they are not stochastic. The decay
characteristic times associated with the modes $2$ and $3$ are calculated as $%
T_{k}=-\left( \ln \left\vert \lambda _{k}\right\vert \right) ^{-1}$,
resulting in the values $T_{2}= 1.000178$ and $T_{3}=0.371878$, so mode $3$
decays faster than mode $2$, while $T_{1}=\infty $ for mode $1$. Thus we can
write 
\begin{equation}
\mathbb{M}^{n}=\mathbb{C}_{1}+\mathbb{C}_{2}\exp \left( -\frac{n}{T_{2}}%
\right) +\mathbb{C}_{3}\exp \left( -\frac{n}{T_{3}}\right) .  \label{C18}
\end{equation}
An initial state vector, for instance, 
\begin{equation}
\mathbb{U}^{\mathrm{T}(0)}=\left( 
\begin{array}{ccc}
100 & 160 & 300%
\end{array}%
\right) ,  \label{C18a}
\end{equation}
will evolve as 
\begin{equation}
\mathbb{U}^{\mathrm{T}}\left( n\right) =\mathbb{U}^{\mathrm{T}}\left(
0\right) \mathbb{C}_{1}+\left( \mathbb{U^{\mathrm{T}}}\left( 0\right) 
\mathbb{C}_{2}\right) \exp \left( -\frac{n}{T_{2}}\right) +\left( \mathbb{U^{%
\mathrm{T}}}\left( 0\right) \mathbb{C}_{3}\right) \exp \left( -\frac{n}{T_{3}%
}\right) .  \label{C19}
\end{equation}
Thereafter, one can follow the evolution of the three components of the
vector (\ref{C19}): 
\begin{subequations}
\label{C20}
\begin{eqnarray}
\mathbb{U}^{\mathrm{T}}\left( 0\right) \mathbb{C}_{1} &=&\left( 
\begin{array}{ccc}
228 & 125 & 207%
\end{array}%
\right) ,  \label{C20a} \\
\mathbb{U^{\mathrm{T}}}\left( 0\right) \mathbb{C}_{2} &=&\left( 
\begin{array}{ccc}
-136 & 110 & -48%
\end{array}%
\right) ,  \label{C20b} \\
\mathbb{U^{\mathrm{T}}}\left( 0\right) \mathbb{C}_{3} &=&\left( 
\begin{array}{ccc}
-67 & -74 & 141%
\end{array}%
\right) ,  \label{C20c}
\end{eqnarray}
\end{subequations}
where $\lim_{n\rightarrow \infty }\mathbb{U}^{\mathrm{T}}\left( n\right) =%
\mathbb{U}^{\mathrm{T}}\left( 0\right) \mathbb{C}_{1} $ is the asymptotic
component and $\mathbb{U^{\mathrm{T}}}\left( 0\right) \mathbb{C}_{l}\times
\exp \left( -n/T_{l}\right) $, $l=2,3$, are the components of the transient
regimes that present negative numbers in some entries, although all the
components of vectors $\mathbb{U}^{\mathrm{T}}\left( n\right) $ and $%
\mathbb{U}^{\mathrm{T}}\left( 0\right) \mathbb{C}_{1} $, Eqs. (\ref{C19})
and (\ref{C20a}), are positive. The sum of the components of $\mathbb{U}^{%
\mathrm{T}}\left( 0\right) \mathbb{C}_{1}$ is $560$ which corresponds to the
conserved sum of the entries of the vector (\ref{C18a}), whereas the sums of
the components of the vectors (\ref{C20b}) and (\ref{C20c}), $%
\sum_{i=1}^{3}\left( \mathbb{U^{\mathrm{T}}}\left( 0\right) \mathbb{C}%
_{2}\right) _{i}=-74$ and $\sum_{i=0}^{3}\left( \mathbb{U^{\mathrm{T}}}%
\left( 0\right) \mathbb{C}_{3}\right) _{i}=0$ respectively, do not need to
be necessarily positive, that we interpret as virtual distributions. As $n$
increases the components $\left( \mathbb{U^{\mathrm{T}}}\left( 0\right) \mathbb{%
C}_{l}\right) \exp \left( -n/T_{l}\right) $ go to zero due to the
exponential decay factor and only the asymptotic vector, Eq. (\ref{C20a}),
survives.
\vspace{2mm}

\textbf{(4)} About the trace operation 
\begin{equation}
\lim_{n\rightarrow \infty }\mathrm{Tr}\left[ \mathbb{M}^{n}\right]
=1+\lim_{n\rightarrow \infty }\sum_{k=2}^{N}\left( \lambda _{k}\right)
^{n}=1.  \label{B1}
\end{equation}
The same holds true for doubly stochastic matrices.
\vspace{2mm}

\textbf{(5)} At the limit $n\rightarrow \infty $ the stationary matrix is 
%
%
\begin{equation}
\mathbb{C}_{1}:=\lim_{n\rightarrow \infty }\mathbb{M}^{n}=
\left( 
\begin{array}{ccccc}
m_{1} & m_{2} & \cdots & m_{N-1} & m_{N} \\ 
m_{1} & m_{2} & \cdots & m_{N-1} & m_{N} \\ 
\vdots & \vdots &  & \vdots & \vdots \\ 
m_{1} & m_{2} & \cdots & m_{N-1} & m_{N} \\ 
m_{1} & m_{2} & \cdots & m_{N-1} & m_{N}%
\end{array}%
\right)  \label{B2}
\end{equation}
\begin{sloppypar}
\noindent with the values of the entries depending on matrix $\mathbb{M}$ and 
$\sum_{j=1}^{N}m_{j}=1$; as all the rows are the same the eigenvalues 
are $\left\{1,0,0,...,0\right\} _{N}$ and the matrix is idempotent,  
$\mathbb{C}_{1}^2 = \mathbb{C}_{1}$. Although the matrix $\mathbb{M}$ 
determines the stationary matrix $\mathbb{C}_{1}$, the inverse is not 
possible, the knowledge of $\mathbb{C}_{1}$ does not permit the full 
determination of $\mathbb{M}$. This is the essence of the irreversible 
evolution. Writing one row of matrix (\ref{B2}) as the vector 
$\mathbb{Z}^{\mathrm{T}} =\left(\begin{array}{ccccc}
m_{1} & m_{2} & \cdots & m_{N-1} & m_{N} \end{array}\right)$, 
it can be noted that it is a stationary distribution because  
$\mathbb{Z}^{\mathrm{T}}\mathbb{M}=\mathbb{Z}^{\mathrm{T}}$ (eigenvalues 1), or 
$\mathbb{C}_{1}\mathbb{M} =\mathbb{C}_{1}$. From Example 1, Eqs. (\ref{C1}) 
and (\ref{C15}), it can be verified that
$ \left( \begin{array}{ccc} 11/27 & 2/9 & 10/27 \end{array} \right) 
\mathbb{M} =\left( \begin{array}{ccc} 11/27 & 2/9 & 10/27 %
\end{array}\right)$. 
\end{sloppypar}

\begin{sloppypar}
A generic vector $\mathbb{X}_{0}^{\mathrm{T}}=\left( \begin{array}{ccccc}
x_{1} & x_{2} & \cdots & x_{N-1} & x_{N}\end{array}\right) $ evolves as 
$\mathbb{X}_{n}^{\mathrm{T}}=\mathbb{X}_{0}^{\mathrm{T}}\mathbb{M}^{n}$  
and asymptotically becomes  
$\mathbb{X}_{\infty }^{\mathrm{T}}=\mathbb{X}_{0}^{\mathrm{T}}\mathbb{C}_{1} $;  
thus it ``looses memory'' about its components but keeps the sum 
$\sum_{i=1}^{N}x_{i}=Y$ as a recollection, i.e., 
$\mathbb{X}_{0}^{\mathrm{T}} \Rightarrow \mathbb{X}_{\infty }^{\mathrm{T}}
=Y\mathbb{Z}^{\mathrm{T}}$. 
\end{sloppypar} 
%
\vspace{2mm}

\textbf{(6)} In the numerical examples to be worked out below the eigenvalues 
of the matrices have multiplicity $1$, nevertheless the math and the physical 
analysis can be extended to the situation where eigenvalues have multiplicity 
higher than $1$. As an illustration about this point we consider here an 
interesting behavior that affects the time decay in the case of occurrence of 
a degenerate eigenvalue (in comparison with the situation of no degeneracy). 
Pointedly, within a closed network, it causes a delay in the time the flux of 
vehicles takes to evolve to the stationary matrix $\lim_{n \rightarrow 
\infty}\mathbb{M}^n$. In this case, the diagonalization of the matrix 
$\mathbb{M}$ -- the decomposition  (\ref{A8}) -- that produces the diagonal 
matrix $\mathbb{D}$ is not anymore possible because there are fewer 
linearly independent eigenvectors than the dimension $N$ of the matrix 
$\mathbb{M}$. Notwithstanding, we can write the decomposition in a form similar 
to (\ref{A8}), $\mathbb{M}=\mathbb{RFR}^{-1}$, which is the so-called 
\emph{Jordan canonical form}, where the matrix $\mathbb{R}$ is different from 
$\mathbb{Q}$, and the matrix $\mathbb{F}$, besides having the eigenvalues 
of $\mathbb{M}$ in the diagonal line, will contain, additionally, in the 
adjacent line parallel to the main diagonal, the number $1$ in the entries 
of the blocks containing the degenerate eigenvalues, while the other entries 
are filled with zeros. The theory is due to the french mathematician Camille 
Jordan. For a rigorous mathematical presentation of the matter we recommend 
the reader to consult, for instance, Ref. \cite{horn2013}. 

The decomposition of a stochastic matrix in the Jordan form can be cast as  
\begin{equation}
\mathbb{M}=\mathbb{RDR}^{-1} + \mathbb{L},\quad \text{and} \quad 
\mathbb{M}^{n}=\mathbb{R} \mathbb{D}^{n}\mathbb{R}^{-1} +f(n)\mathbb{L},  
\label{A12}
\end{equation}
where $\mathbb{D}$ is still a diagonal matrix containing the eigenvalues, 
$\mathbb{L}=\mathbb{R}\mathbb{O}\mathbb{R}^{-1}$ with $\mathbb{O}$ the matrix 
containing $1$'s (and $0$'s) in a secondary diagonal line of matrix $\mathbb{F}$ 
and $f(n)$ is a function only of $n$. We illustrate this case through an example.

\vspace{2mm} 
\noindent \textbf{\underline{Example 2:}} Let us consider the doubly stochastic matrix
\begin{eqnarray}
\mathbb{M} &=&\left( 
\begin{array}{ccc}
2/5 & 1/2 & 1/10 \\
3/10 & 3/10 & 2/5 \\
3/10 & 1/5 & 1/2%
\end{array}%
\right) ;\quad \det \mathbb{M} = 1/100\ , \label{A133} 
\end{eqnarray}%
whose eigenvectors and eigenvalues are
\begin{equation}
\left( 
\begin{array}{c}
1 \\ 
1 \\ 
1%
\end{array}%
\right) \leftrightarrow \chi_{PF} = 1,\quad \left( \begin{array}{r}
-2 \\ 
1 \\ 
1%
\end{array}%
\right) \leftrightarrow \chi_{2}=\frac{1}{10},\  \text{multiplicity 2}\ ,
\label{A14}
\end{equation}
$\chi_{PF}$ is the Perron-Frobenius eigenvalue and $\chi_{2}$ is 
degenerate, with only one linearly independent eigenvector. The 
decomposition of matrix (\ref{A133}) in the Jordan form is
\begin{equation}
\mathbb{M} =\mathbb{RFR}^{-1}=\left( 
\begin{array}{rrr}
1 & 0 & 1 \\ 
1 & 5/2 & -1/2 \\ 
1 & -5/2 & -1/2%
\end{array}%
\right) \left( 
\begin{array}{rrr}
1 & 0 & 0 \\ 
0 & \chi_2 & 0 \\ 
0 & 1 & \chi_2%
\end{array}%
\right) \left( 
\begin{array}{rrr}
1/3 & 1/3 & 1/3 \\ 
0 & 1/5 & -1/5 \\ 
2/3 & -1/3 & -1/3%
\end{array}%
\right) 
\label{A15}
\end{equation}
where the matrix $\mathbb{F}$ is not diagonal although the eigenvalues 
remain in the main diagonal. We can rewrite the matrix (\ref{A15}) as
\begin{equation}
\mathbb{M} = \mathbb{RDR}^{-1}+f\left( 1\right) \mathbb{L}
= \mathbb{C}_{1}\allowbreak +\chi _{2}\mathbb{C}_{2} +\chi
_{2}\mathbb{C}_{3}
\label{A16}
\end{equation}
where the matrices $\mathbb{C}_{i}$ are
\begin{equation}
\mathbb{C}_{1}=\frac{1}{3}\left( 
\begin{array}{ccc}
1 & 1 & 1 \\ 
1 & 1 & 1 \\ 
1 & 1 & 1%
\end{array}%
\right) ,\quad \mathbb{C}_{2}=\left( 
\begin{array}{rrr}
2/3 & -1/3 & -1/3 \\ 
-1/3 & 2/3 & -1/3 \\ 
-1/3 & -1/3 & 2/3%
\end{array}%
\right) ,\quad \mathbb{C}_{3}=\left( 
\begin{array}{rrr}
0 & 2 & -2 \\ 
0 & -1 & 1 \\ 
0 & -1 & 1%
\end{array}%
\right) \ .
\label{A17}
\end{equation}
For an $n$-step process we have
$
\mathbb{M}^{n}=\mathbb{C}_{1}+\left( \chi _{2}\right) ^{n}\mathbb{C}%
_{2}+n\left( \chi _{2}\right) ^{n}\mathbb{C}_{3}.$ 
Setting $\chi_{2} = \exp \left( -1/T_2 \right) $ we get
\begin{equation}
\mathbb{M}^{n}=\mathbb{C}_{1}+\exp \left({-\frac{n}{T_{2}}}\right)\mathbb{C}_{2}
+ n\exp\left({-\frac{n}{T_{2}}}\right)  \mathbb{C}_{3}
\label{A18}
\end{equation}
with the characteristic decay time $T_{2}=-\left( \ln \chi _{2}\right)
^{-1}=\left( \ln 10\right) ^{-1}\approx  0.43$ for the exponential decay modes. 
Due to the factor $n$ that multiplies the exponential in the third term, it is 
going to decay at a lower pace than the second one, thus dominating the 
evolution of $\mathbb{M}^{n}$ toward the stationary matrix 
$\mathbb{C}_{1}$ as $n\rightarrow \infty$.
%
%
%
%
%
\subsection{Entropy}
%
The complexity of the flow of vehicles in a network, as the one represented
by the digraph (\ref{fig01}), has a measure that can be evaluated by
calculating the entropy at each moment $n$. Since $%
\sum_{j=1}^{N}\left( \mathbb{M}^{n}\right) _{ij}=1$ ($N$ stands for the
dimension of the square matrix), for the SM we define
the partial entropy associated to each row $i$ as 
\begin{equation}
S_{i}\left( n\right) =-\sum_{j=1}^{N}\left( \mathbb{M}^{n}\right) _{i,j}\ln
\left( \mathbb{M}^{n}\right) _{i,j}\qquad \mathrm{for\quad }i=1,...,N;
\label{Ent1}
\end{equation}
noting that if no vehicle migrates from one site to the other, then $\left( 
\mathbb{M}^{n}\right) _{ij}=\delta _{ij}$ and $S_{i}\left( n\right) =0$. We
can also define a measure of the global complexity as an 
arithmetic mean over all the sites, 
\begin{equation}
G\left( n\right) =\frac{1}{N}\sum_{i=1}^{N}S_{i}\left( n\right) =-\frac{1}{N}%
\sum_{i=1}^{N}\sum_{j=1}^{N}\left( \mathbb{M}^{n}\right) _{ij}\ln \left( 
\mathbb{M}^{n}\right) _{ij}\ ,  \label{Ent2}
\end{equation}
outlining a \emph{global mean entropy}.
%
\section{Model I: Illustration with numbers}
%
In the four sites network (digraph of Fig. \ref{fig01}) we assume that 
there are, for instance, initially, $10\ 000$, $2\ 500$, $11\ 000$ and 
$13\ 000$ vehicles; the matrix (\ref{A1}) with numbers in its entries is 
diagonal and is more conveniently expressed as the vector (\ref{A22}),
\begin{equation}
\mathbb{U}^{\mathrm{T}}\left( 0\right) =\left( 
\begin{array}{cccc}
10000 & 2500 & 11000 & 13000%
\end{array}%
\right) ,  \label{B02}
\end{equation}
and the total number of vehicles in the network is 
\begin{equation}
Y=\sum_{i=1}^{4}U_{i}\left( 0\right) =36\ 500.  \label{B1b}
\end{equation}

At a given moment the vehicles begin to circulate through the arteries 
and after few days (or at several moments of one day) of observation the 
average number of the vehicles that remain circulating (or parked) within 
each site (the vertices) are assumed \emph{ad hoc} to be
$3\ 000$, $1\ 000$, $4\ 000$, and $4\ 500$, while the remaining vehicles are on
their way traveling to the other sites, characterized in the digraph by the
values associated to the edges. We put the available information in a
matrix form 
\begin{equation}
\mathbb{A}\left( 1\right) =\left( 
\begin{array}{rrrr}
3000 & 1500 & 2500 & 3000 \\ 
500 & 1000 & 500 & 500 \\ 
3000 & 1500 & 4000 & 2500 \\ 
4000 & 1500 & 3000 & 4500%
\end{array}%
\right) \ , \label{B0}
\end{equation}
where in the diagonal entries one finds the number of vehicles in each site and the
off-diagonal entries stand for the number of those traveling from one site
(line $i$) to another (column $j$), and \textrm{Tr}$\left( \mathbb{A}\left(
1\right) \right) = 12\ 500$ is the number of vehicles still at the sites. The
number of vehicles in each row ($U_{k}\left( 0\right) $) are given by the vector (\ref%
{B02}). %
%
\subsection{The Stochastic Matrix}
%
We assume that the matrix (\ref{B0}) is the seed that permits to construct another 
matrix, in a suited form, to be used to forecast the circulation of the vehicles 
in future moments. This hypothesis is framed formally by normalizing each row. 
The result is the SM 
\begin{equation}
\mathbb{M}=\left( 
\begin{array}{cccc}
3/10 & 3/20 & 1/4 & 3/10 \\ 
1/5 & 2/5 & 1/5 & 1/5 \\ 
3/11 & 3/22 & 4/11 & 5/22 \\ 
4/13 & 3/26 & 3/13 & 9/26%
\end{array}%
\right)\ ,  \label{D1}
\end{equation}
whose eigenvalues are approximately
\begin{equation}
\begin{tabular}{|c||c|c|c|c|}
\hline
$k$ & $1$ & $2$ & $3$ & $4$ \\ \hline
\multicolumn{1}{|l||}{$\lambda _{k}$} & \multicolumn{1}{l|}{${\ 1.0}$} & 
\multicolumn{1}{l|}{${\ 0.27}$} & \multicolumn{1}{l|}{${\ 0.13}$} & 
\multicolumn{1}{l|}{${\ 0.01}$} \\ \hline
\end{tabular}%
\ .  \label{D1.1}
\end{equation}
The eigenvalue $\lambda _{1}=1.0$ is the PF while the others are
quite smaller than $1$, this is an indication that under this kind of 
evolution the traffic should stabilize after very few steps. As $\det \mathbb{%
M}=\left({2600}\right)^{-1}<1$, then $\lim_{n\rightarrow \infty }$\bigskip $\det \left( 
\mathbb{M}^{n}\right) =0$ and $\lim_{n\rightarrow \infty }\mathrm{Tr}\left( 
\mathbb{M}^{n}\right) =1$.

\begin{figure}[H]
\centering
\includegraphics[height=2.94in, width=3.94in]{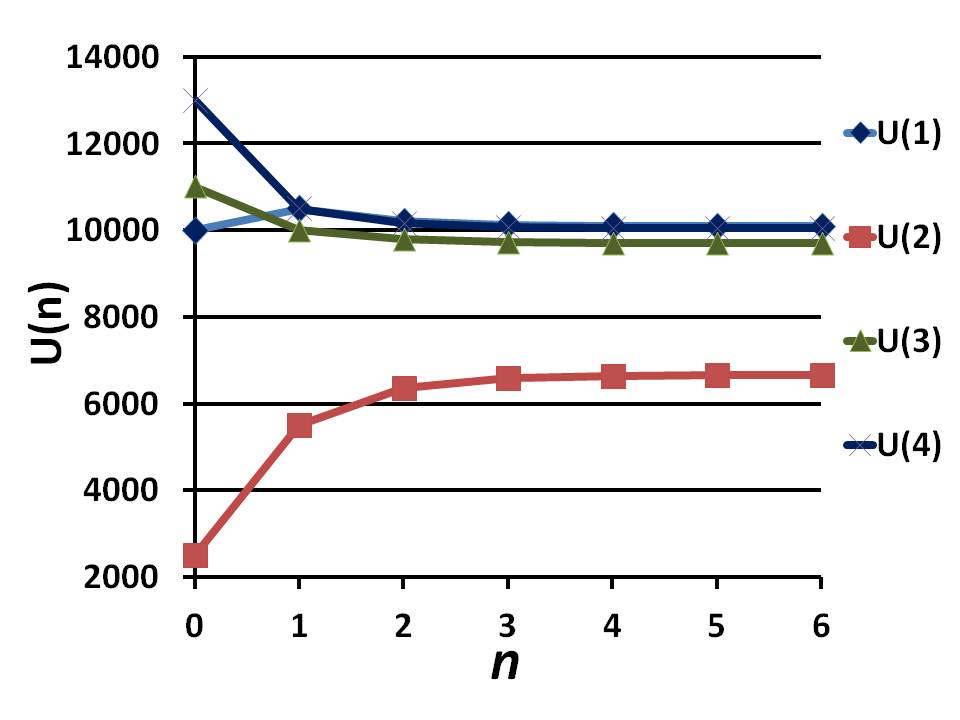}
\caption{{\small {The sequences of the distributions $U_{i}(n)$.}}}
\label{fig031}
\end{figure}

Now we calculate the number of vehicles in each site plus those that left (in a
period of 24 hours). The prediction of the distribution for the following days 
is given by Eq. (\ref{A24}), and the total number of vehicles is a conserved 
quantity, $\lim_{n\rightarrow \infty }\sum_{i}U_{i}\left( n\right) =36\,500 = Y$. 
The distribution changes with $n$ until the stabilization of the flow is attained 
swiftly in less than $4$ steps. This trend could be guessed due to the wide difference
between the PF eigenvalue ($\lambda _{1}$=1) and $\lambda _{2}$, see the frame (%
\ref{D1.1}). The vehicular circulation presents three characteristic decay 
times towards the stationary distribution 
\begin{equation}
\begin{tabular}{|c||c|c|c|c|}
\hline
$i$ & ${\ 1}$ & ${\ 2}$ & ${\ 3}$ & ${\ 4}$ \\ \hline
\multicolumn{1}{|l||}{$T_{i}$} & \multicolumn{1}{l|}{${\ \infty }$} & 
\multicolumn{1}{l|}{${\ 0.76}$} & \multicolumn{1}{l|}{${\ 0.49}$}
& \multicolumn{1}{l|}{${\ 0.22}$} \\ \hline
\end{tabular}\ ,  
\label{C3}
\end{equation}
where $T_2$ dictates the trend of the decay. In Fig. \ref{fig031} we plot 
the sequences of the distributions $U_{i}^{\mathrm{T}}(n)$ that stabilize 
swiftly to asymptotic ones. For $\mathbb{U}^{\mathrm{T}}{\ (0)}=\left( 
\begin{array}{cccc}
{10000} & {\ 2500} & {\ 11000} & {\ 13000}%
\end{array}
\right) $ we get $\mathbb{U}^{\mathrm{T}}{\ (\infty )
\approx \mathbb{U}^{\mathrm{T}}{\ (6)} =}\left( \begin{array}{cccc}
{10097} & {\ 6659} & {\ 9702} & {\ 10042}%
\end{array}
\right) $. 
%
\subsection{Entropy}
\begin{figure}[H]
\centering
\includegraphics[height=3.0in, width=4.2in]{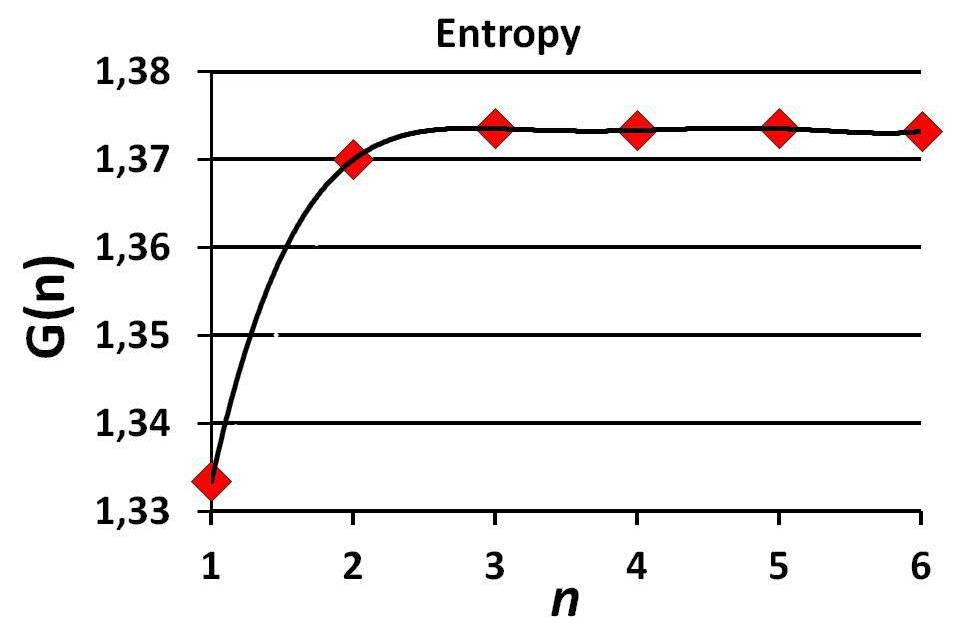}
\caption{{\small {The global mean entropy as defined in Eq. (\ref{Ent2}).}}}
\label{fig010}
\end{figure}

%
The values of the global mean entropy, (\ref{Ent2}), are given in 
Table \ref{D4}, 
\begin{table}[H]
\centering
\begin{tabular}{|c||c|c|c|c|c|c|c|c|}
\hline
${\ n}$ & $1$ & $2$ & $3$ & $4$ & $5$ & $6$ \\ \hline
\multicolumn{1}{|l||}{${\ G}\left( n\right) $} & \multicolumn{1}{l|}{${\
1.333}$} & \multicolumn{1}{l|}{${\ 1.372}$} & \multicolumn{1}{l|}{${\ 1.373}$%
} & \multicolumn{1}{l|}{${\ 1.373} $} & \multicolumn{1}{l|}{${\ 1.373}$} & 
\multicolumn{1}{l|}{${\ 1.373}$}   
 \\ \hline
\end{tabular}%
\caption{{\small {Numerical values for the global mean entropy, Eq. (\ref{Ent2}).}}}
\label{D4}
\end{table}
\noindent and in Fig. \ref{fig010} the diamond shaped marks represent the 
calculated values of $G\left( n\right) $ of Table \ref{D4}; the solid line 
only links the points. 
The stabilization of the distribution of vehicles in each site, and the
stationarity of the flow is reflected in the behavior of the global mean
entropy which increases swiftly and then stabilizes at a maximum value, around 
$1.37$, that we could call the ``thermalization'' of the traffic. 
%
\section{Model II: Inter-site traffic with input (source) and 
output (sink) of vehicles}
\begin{figure}[htbp]
\centering
\includegraphics[height=2.44in, width=2.6in]{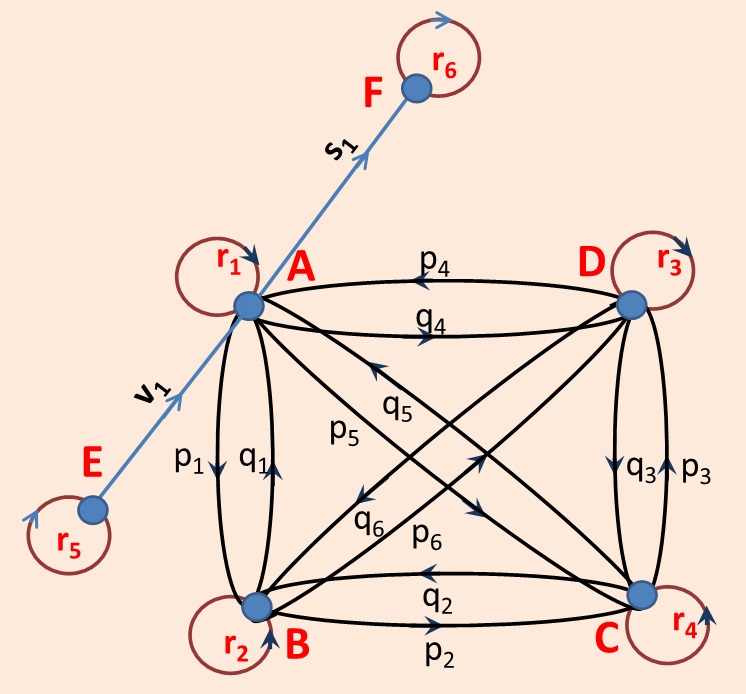}
\caption{{\small{Two physical modifications in the inter-site
traffic as drawn in Fig. \ref{fig01}}}}
\label{fig03}
\end{figure}

%
We now introduce two physical modifications in the inter-site
traffic model studied in the preceding sections: (a) there is a daily input
of new vehicles from site $E$ to city $A$, that are integrated into the
network and (b) there is also an output of old vehicles that are taken off
the circuit from the same site $A$ and are stocked in the site $F$. These new
features are drawn in the digraph with additional edges and loops, labeled
as $v_{1}$ and $v_{2}$, $r_{5}$ and $r_{6},$ respectively, as shown in Fig. %
\ref{fig03}. 
In this digraph the letters $A$, $B$, $C$, $D$, $E$ and $F$ label the 
vertices, whereas the $r_{i}$'s are for the loops and $q_{i}$, $p_{i}$, 
$s_1$ and $v_1$ are for the edges. One expresses the network by the array 
in Table \ref{A5}, 
\begin{table}[htbp]
\centering
\begin{tabular}{|c||l|l|l|l|l|l||l|}
\hline
Vertices & $A$ & $B$ & $C$ & $D$ & $E$ & $F$ & Sum of the line entries \\ 
\hline\hline
$A$ & $R_{1}$ & $P_{1}$ & $P_{5}$ & $Q_{4}$ & $0$ & $S_{1}$ & 
\multicolumn{1}{c|}{${\ W}_{1}$} \\ \hline
$B$ & $Q_{1}$ & $R_{2}$ & $P_{2}$ & $P_{6}$ & $0$ & $0$ & 
\multicolumn{1}{c|}{${\ W}_{2}$} \\ \hline
$C$ & $Q_{5}$ & $Q_{2}$ & $R_{3}$ & $P_{3}$ & $0$ & $0$ & 
\multicolumn{1}{c|}{${\ W}_{3}$} \\ \hline
$D$ & $P_{4}$ & $Q_{6}$ & $Q_{3}$ & $R_{4}$ & $0$ & $0$ & 
\multicolumn{1}{c|}{${\ W}_{4}$} \\ \hline
$E$ & ${\ V}_{1}$ & $0$ & $0$ & $0$ & $R_{5}$ & $0$ & \multicolumn{1}{c|}{${%
\ W}_{5}$} \\ \hline
$F$ & $0$ & $0$ & $0$ & $0$ & $0$ & $R_{6}$ & \multicolumn{1}{c|}{${\ W}_{6}$%
} \\ \hline\hline 
&  &  &  &  &  &  & \multicolumn{1}{c|}{} \\
Sum of the column entries & ${\ Z}_{1}$ & ${\ Z}_{2}$ & ${\ Z}_{3}$ & ${\ Z}%
_{4}$ & ${\ Z}_{5}$ & ${\ Z}_{6}$ & \multicolumn{1}{c|}{$\overline{Y}$} \\ 
\hline
\end{tabular}%
\caption{{\small {The labels $A$, $B$, $C$, $D$, $E$ and $F$ specify
the vertices in the digraph, the entries $Q_i$, $P_i$, $V_1$ and $S_1$ stand
for the numbers of vehicles trafficking along the roads, from one site 
to the other, and the $R_i$'s are the numbers of vehicles present at each
site. The sum of the entries of each row and column are $W_i$ and $Z_i$
respectively. $\overline{Y}=\sum_{i=1}^{6}W_{i} = \sum_{i=1}^{6}Z_{i}$ is
the total number of vehicles in the network.}}}
\label{A5}
\end{table}
whose core is the matrix (\ref{A6})
\begin{equation}
\mathbb{B}=\left( 
\begin{array}{cccccc}
R_{1} & P_{1} & P_{5} & Q_{4} & 0 & S_{1} \\ 
Q_{1} & R_{2} & P_{2} & P_{6} & 0 & 0 \\ 
Q_{5} & Q_{2} & R_{3} & P_{3} & 0 & 0 \\ 
P_{4} & Q_{6} & Q_{3} & R_{4} & 0 & 0 \\ 
V_{1} & 0 & 0 & 0 & R_{5} & 0 \\ 
0 & 0 & 0 & 0 & 0 & R_{6}%
\end{array}%
\right) ,  \label{A6}
\end{equation}
whose entries contain the number of counted vehicles in an \emph{ad hoc} 
interval of time, or by averaging from previous observations. The matrix 
(\ref{A1}) is a submatrix of matrix (\ref{A6}).

As the time goes on the distribution of vehicles, in matrix (\ref{A6}), changes 
due to their circulation. The sum of the entries of row $i$, $W_{i}\left( 0\right)
=\sum_{j}B_{ij}$, gives the number of vehicles that are in site $i$ plus those
that left it having as destination all the other sites, excluding site $F$, 
which is a depository of the vehicles removed from circulation. The total number of 
vehicles in the network, $\overline{Y}$, is conserved. The numbers 
$W_{i}\left( 0\right) $ can be cast as a vector,
\begin{equation}
\mathbb{W}^{\mathrm{T}}\left( 0\right) =\left( 
\begin{array}{cccccc}
W_{1}\left( 0\right) & W_{2}\left( 0\right) & W_{3}\left( 0\right) & 
W_{4}\left( 0\right) & W_{5}\left( 0\right) & W_{6}\left( 0\right)%
\end{array}%
\right) .
\end{equation}
%
\subsection{The stochastic matrix}
%
The dynamical evolution is ruled by a stochastic matrix associated to (\ref{A6}) 
which is constructed by normalizing the entries in each row, resulting in 
\begin{equation}
\mathbb{N}=\left( 
\begin{array}{cccccc}
r_{1} & p_{1} & p_{5} & q_{4} & 0 & s_{1} \\ 
q_{1} & r_{2} & p_{2} & p_{6} & 0 & 0 \\ 
q_{5} & q_{2} & r_{3} & p_{3} & 0 & 0 \\ 
p_{4} & q_{6} & q_{3} & r_{4} & 0 & 0 \\ 
v_{1} & 0 & 0 & 0 & r_{5} & 0 \\ 
0 & 0 & 0 & 0 & 0 & 1%
\end{array}%
\right) \quad ,  \label{A6a}
\end{equation}
where the first row is 
\begin{equation}
r_{1}=\frac{R_{1}}{W_{1}},\quad p_{1}=\frac{P_{1}}{W_{1}},\quad p_{5}=\frac{%
P_{5}}{W_{1}},\quad q_{4}=\frac{Q_{4}}{W_{1}},\quad s_{1}=\frac{S_{1}}{W_{1}}%
;  \label{A6b}
\end{equation}
and the same goes for the other rows. The evolution of an initial vector is 
calculated as 
\begin{equation}
\mathbb{W}^{\mathrm{T}}\left( n\right) =\mathbb{W}^{\mathrm{T}}\left(
0\right) \mathbb{N}^{n}.  \label{A6c}
\end{equation}
%
\section{Model II: Illustrating the model with numbers}
%
To illustrate the model we consider the matrix (\ref{A6a}) 
\begin{equation}
\mathbb{B}\left( 1\right) =\left( 
\begin{array}{cccccc}
{\ 3000} & {\ 1500} & {\ 2500} & {\ 3000} & {\ 0} & {\ 50} \\ 
{\ 500} & {\ 1000} & {\ 500} & {\ 500} & {\ 0} & {\ 0} \\ 
{\ 3000} & {\ 1500} & {\ 4000} & {\ 2500} & {\ 0} & {\ 0} \\ 
{\ 4000} & {\ 1500} & {\ 3000} & {\ 4500} & {\ 0} & {\ 0} \\ 
{\ 100} & {\ 0} & {\ 0} & {\ 0} & {\ 1000} & {\ 0} \\ 
{\ 0} & {\ 0} & {\ 0} & {\ 0} & {\ 0} & {\ R}_{6}%
\end{array}%
\right) ,  \label{A7}
\end{equation}
where we kept the same values of the entries of the $4\times 4$ matrix (\ref%
{B0}) but being now enlarged, with dimensions $6\times 6$, and with new non-null 
entries, $B_{16}=50$, $B_{51}=100$, $B_{55}=1000$; the entry $B_{66}=R_{6}$ is 
an arbitrary number, that we set as $0$ since it is not important within the 
dynamics. Summing the entries in each row of matrix (\ref{A7}) we have the vector 
\begin{equation}
\mathbb{W}^{\mathrm{T}}\left( 0\right) =\left( 
\begin{array}{cccccc}
{\ 10050} & {\ 2500} & {\ 11000} & {\ 13000} & {\ 1100} &0 %
\end{array}%
\right)  \label{A9}
\end{equation}
or as a diagonal matrix $\mathbb{B}\left( 0\right) = \mathrm{Diag}%
[10050,2500,11000,13000,1100,0]$ and the total number of vehicles is 
$\overline{Y}=37\, 650$. Comparing with Model I, the increment in the 
number of vehicles in vector (\ref{A9}) is small relatively to those 
in the string (\ref{B02}), namely an increment of $1\, 150$ to the 
previous $36\, 500$. 
%
\subsection{The Stochastic Matrix}
%
From the matrix (\ref{A7}) we construct the stochastic matrix 
\begin{equation}
\mathbb{N}=\left( 
\begin{array}{cccccc}
300/1005 & 150/1005 & 250/1005 & 300/1005 & 0
& 5/1005 \\ 
1/5 & 2/5 & 1/5 & 1/5 & 0 & 0 \\ 
3/11 & 3/22 & 4/11 & 5/22 & 0 & 0 \\ 
4/13 & 3/26 & 3/13 & 9/26 & 0 & 0 \\ 
1/11 & 0 & 0 & 0 & 10/11 & 0 \\ 
0 & 0 & 0 & 0 & 0 & 1%
\end{array}%
\right)\quad ,  \label{A11}
\end{equation}
whose eigenvalues are, approximately, 
\begin{equation}
\begin{tabular}{|c||c|c|c|c|c|c|}
\hline
${\ k}$ & ${\ 1}$ & ${\ 2}$ & ${\ 3}$ & ${\ 4}$ & ${\ 5}$ & ${\ 6}$ \\ \hline
${\ \lambda }_{k}$ & ${\ 1.0}$ & ${\ 0.999}$ & ${\ 0.909}$ & ${\
0.266}$ & ${\ 0.132}$ & ${\ 0.011}$ \\ \hline
\end{tabular}%
\ .  \label{A11.1}
\end{equation}
The first eigenvalue is the PF while the second and third ones are close to $%
1$, making the traffic to take much more time to attain a stationary circulation 
than in the former model where the traffic goes stationary in 3 or 4 steps; 
these eigenvalues can be compared to those in frame (\ref{D1.1}).

\begin{figure}[htbp]
\centering
\includegraphics[height=3.0in, width=4.0in]{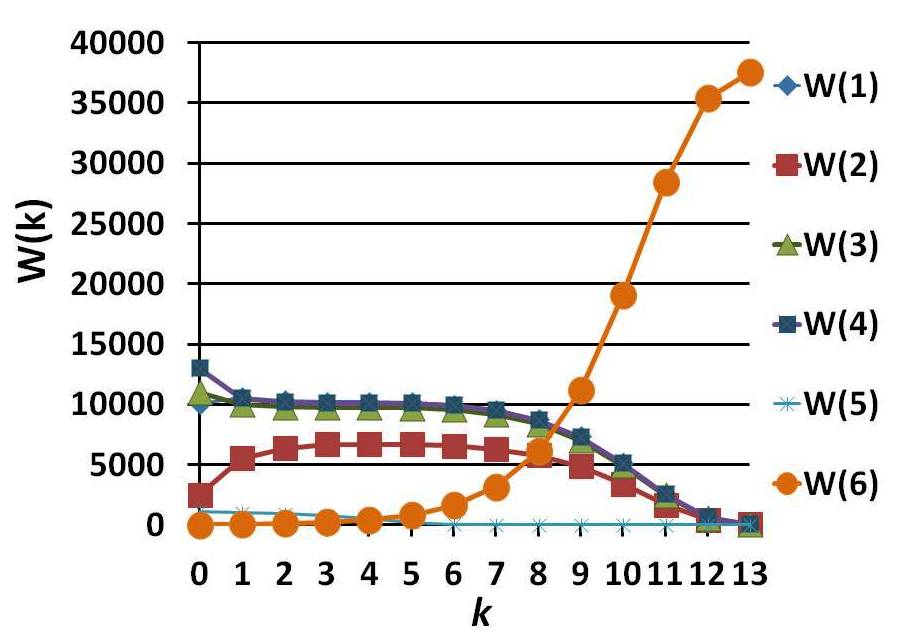}
\caption{{\small {The evolution of the number of vehicles in each component
of the vector $\mathbb{W}\left( n\right) $, at several moments, where we
defined $k = 1+\log _{2}n$ as the abscissa variable. We have set \emph{ad hoc}
$k = 0 $ for $n = 0$.}}}
\label{fig04}
\end{figure}

\begin{figure}[htbp]
\centering
\includegraphics[height=2.7in, width=3.6in]{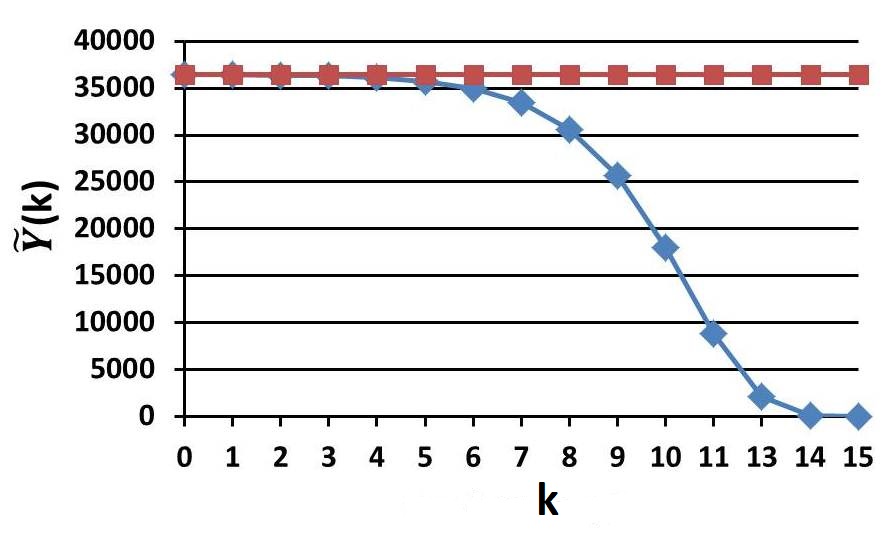}
\caption{{\small {The line in blue (diamonds marks) is the evolution
of the number of vehicles within the network $ABCD$, $\tilde{Y} \left( n\right) $, 
with $k=\log_{2} n +1$ and for $n=0$ we have set $k=0$. 
The red line (square marks) stands for the network without source and sink.}}}
\label{fig08}
\end{figure}

The distribution of vehicles evolves as $\mathbb{W}^{\mathrm{T}}\left(
n\right) =\mathbb{W}^{\mathrm{T}}\left( 0\right) \mathbb{N} ^{n}$, 
with $\mathbb{W}^{\mathrm{T}}\left( 0\right)$ from Eq. (\ref{A9}). For
$n=2^{12}$ we get $\mathbb{W}^{\mathrm{T}}\left( 2^{12}\right) =\left( 
\begin{array}{cccccc} {\ 37} & {\ 24} & {\ 36} & {\ 37} & {\ 0} & {37516}%
\end{array} \right) $. The last entry stands for the $37\ 516$ vehicles that 
went out of circulation, at an average daily fraction $1/201$ of all vehicles, 
stockpiled in site $F$, whereas the number of vehicles still in circulation 
diminishes continuously. The fifth entry stands for the supply of new 
vehicles into the network: there were initially $1\ 000$ vehicles at site 
$E$, with a daily supply into the network $ABCD$ of $1/11$ fraction of 
vehicles still in the yard $E$ (the supply lasted few days). We note that 
after $2^{12}$ steps only a tiny fraction of vehicles (134), $0.36\%$, 
continue circulating within the network $ABCD$. In Fig. \ref{fig04} we plot 
the evolution of the entries of $\mathbb{W}\left( n\right)$, considering 
in the abscissa $k\equiv \left( \log _{2}n\right) +1$, for 
$n=2^0,2^1,...,2^{12}$, such that $k=1,2,..,13$. 
Thus, after an initial increase of the number of vehicles circulating within the
network $ABCD$ there is a stabilization, see Fig. \ref{fig08}, then begins a slow 
decrease in their number, however after consuming a quite long time compared to the 
time it takes to attain equilibration in the model without source and sink. 

We note that for the digraph in Fig. \ref{fig01} the asymptotic equilibrium
of the evolution is attained nearly after four steps, $n=4$, whereas for the 
network with source and sink, the digraph in Fig. \ref{fig03}, the tendency
to the stationary distribution, $\mathbb{W}\left( 2^{12} \right)$, occurs after 
$2^{12}$ steps, or for $\left( \mathbb{N}\right) ^{2^{12}}$. In short, in the former 
network model the stationarity is reached at a rate linear in time, while here  
it occurs at an exponential rate. The characteristic decay times 
$\tilde{T}_{k}=-\left( \ln \left\vert \lambda _{k}\right\vert \right) ^{-1}$ are 
\begin{equation}
\begin{tabular}{|c||c|c|c|c|c|c|}
\hline
${\ k}$ & ${\ 1}$ & ${\ 2}$ & ${\ 3}$ & ${\ 4}$ & ${\ 5}$ & ${\ 6}$ \\ \hline
&  &  &  &  &  &  \\ 
${\ \tilde{T}}_{k}$ & ${\ \infty }$ & \multicolumn{1}{l|}{${\ 726.20}$} & 
\multicolumn{1}{l|}{${\ 10.50}$} & \multicolumn{1}{l|}{${\ 0.76}$} & 
\multicolumn{1}{l|}{${\ 0.49}$} & \multicolumn{1}{l|}{${\ 0.22}$}
\\ \hline
\end{tabular}%
~,  \label{A13}
\end{equation}
to be compared to the characteristic times in the frame (\ref{C3}). 
Since $\tilde{T}_{2}/T_{2}\approx 961$, $\tilde{T}_{3}/T_{3}\approx
\allowbreak 13.9$, $\tilde{T}_{4}/T_{4}\approx 3.4$, we perceive the 
huge ratio of the characteristic times $\tilde{T}_{2}/T_{2}$ 
meaning that the introduction of a source and a sink in the $ABCD$ network, 
for the specific chosen numbers, contribute to an average time (to attain 
the stationary state) that is about three orders of magnitude higher than 
the network without source and sink. 
%
\subsection{Entropy}
\begin{figure}[htbp]
\centering
\includegraphics[height=2.6in, width=3.4in]{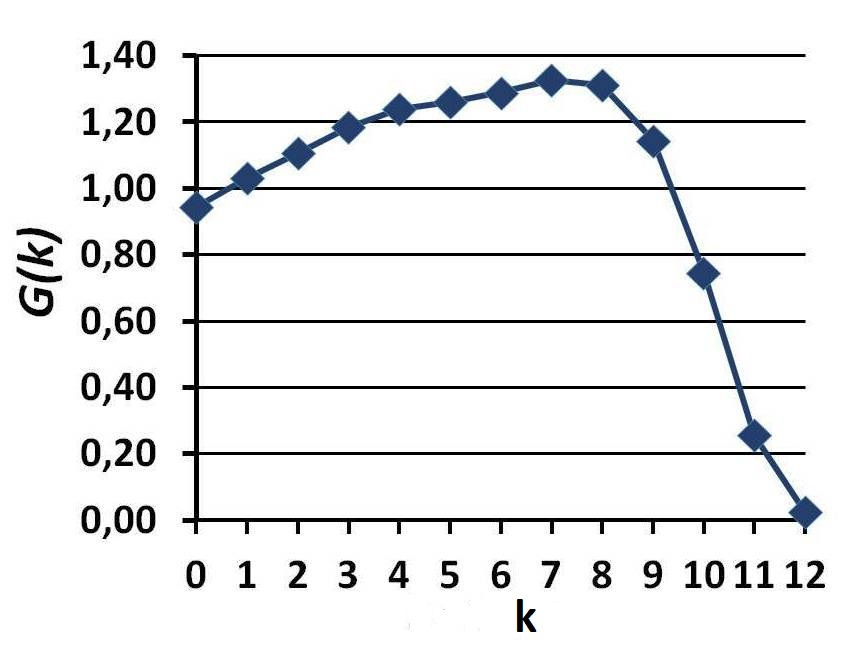}
\caption{{\small {Global mean entropy ${G}\left( k\right) $ for the
stochastic matrix $\mathbb{N}^{n}$, where $k=\log _{2}n$.}}}
\label{fig06}
\end{figure}

%
The global mean entropy, Eq. (\ref{Ent2}), for $N=6$, at times $n=2^{k}$ 
with $k=0,1,2,...,12$ is plotted in Fig. \ref{fig06}. Compared to Fig. \ref{fig010}  
it shows a quite different trend, the entropy increases, attains a maximum value 
and then begins a monotonic decrease until zero, when all the vehicles 
accumulate at the site $F$, as they are being continuously collected. 
%
\section{Nonlinear model: multiple stationary states and cyclic changes}
%
More realistically one cannot expect that the traffic flow could be described
strictly by a linear model, even with sources and sinks in the network, since it 
leads to only one stationary state. Thus adopting a nonlinear approach  
seems more realistic because it introduces non-trivial changes. In this model 
the entries of a stochastic matrix contain functions of $n$ as an intrinsic 
variable, and there is a freedom to choose functions and parameters in order 
to emulate a flow displaying several stationary regimes in one single day, 
for instance.

For the sake of illustration we consider a two-site 
network ($A$ and $B$) and two arteries connecting them, with no sinks or sources. 
We assume that the stochastic matrix has dimension 2 and we write it as 
\begin{equation}
 \mathbb{M} \left( n \right)  =\left( 
\begin{array}{cc}
g_{1}\left( n\right) &  1-g_{1}\left( n\right) \\ \\
g_{2}\left( n\right) & \ 1-g_{2}\left( n\right)%
\end{array}%
\right) \ , 
\label{L1}
\end{equation}
whose entries are $n$-dependent and remain non-negative for any positive 
integer $n$, as long as $0 \leq g_{k}\left( n\right) \leq 1$. 
Considering that the functions $g_{k}\left( n\right) $ change periodically 
with $n$, we set $g_{1}\left( n\right) = a_1 + b_1\sin \left( \frac{2\pi n}{L}\right) $ 
and $g_{2}\left( n\right) = a_2 + b_2\cos \left( \frac{2\pi n}{L}\right) $, 
with $a_1=a_2=1/2$, $b_1=b_2=-1/2$, and, as it will be seen below, the integer 
$L$ stands for the multiplicity of the stationary states. For the sake 
of illustration we adopt $L=3$, which implies three asymptotic matrices 
of an $n$-step evolution. This property for nonlinear stochastic matrices 
enlarges the descriptive possibilities of physical systems in comparison with 
the linear models. With those settings the   
matrix (\ref{L1}) is written as 
\begin{equation}
\mathbb{M} \left( n \right) =\left( 
\begin{array}{cc}
\frac{1}{2}\left( 1-\sin \frac{2\pi n}{3}\right) & \frac{1}{2}\left( 1+\sin 
\frac{2\pi n}{3}\right) \\ \\
\frac{1}{2}\left( 1-\cos \frac{2\pi n}{3}\right) & \frac{1}{2}\left( 1+\cos 
\frac{2\pi n}{3}\right)%
\end{array}%
\right)\ ,  \label{L2}
\end{equation}
and is 3-cycle, meaning that for each element of the set $n=\{3k,3k+1,3k+2\}$ 
(excluding $n=0$) the matrix $\mathbb{M}\left( n\right)$ repeats itself, as 
shown in the Table \ref{Tnl1}. 
\begin{table}[H]
\centering
\begin{tabular}{|c||c|c|c|}
\hline
$k$ & $n=3k$ & $n=3k+1$ & $n=3k+2$ \\ \hline\hline
$0$ & $0$ & $1$ & $2$ \\ \hline
$1$ & $3$ & $4$ & $5$ \\ \hline
$2$ & $6$ & $7$ & $8$ \\ \hline
$\vdots $ & $\vdots $ & $\vdots $ & $\vdots $ \\ \hline
\end{tabular}%
\caption{{\small{The integer $k$ maps one-to-one the sequence of integers $n$.}}}
\label{Tnl1}
\end{table}

\noindent The stochastic matrices and decay modes eigenstates and eigenvalues 
are given in Table \ref{Tnl2}, and their expansion in the several modes and 
their decay times are given in Table \ref{Tnl3}.
\begin{table}[H]
\centering
\begin{tabular}{|c|c|}
\hline
$n$-step stochastic matrices & Decay mode eigenstate and eigenvalue \\ 
\hline\hline
$\mathbb{M}\left( 3k+1\right) \doteq \mathbb{M}_{1}=\left( 
\begin{array}{cc}
\frac{1}{2}-\frac{\sqrt{3}}{4} & \frac{1}{2}+\frac{\sqrt{3}}{4} \\ 
\frac{3}{4} & \frac{1}{4}%
\end{array}%
\right) $ & $\frac{1}{2\sqrt{4+\sqrt{3}}}\left( 
\begin{array}{c}
-\left( \sqrt{3}+2\right) \\ 
3%
\end{array}%
\right) \leftrightarrow -\frac{1}{4}\left( \sqrt{3}+1\right) =\lambda _{1}$
\\ \hline
$\mathbb{M}\left( 3k+2\right) \doteq \mathbb{M}_{2}=\left( 
\begin{array}{cc}
\frac{1}{2}+\frac{\sqrt{3}}{4} & \frac{1}{2}-\frac{\sqrt{3}}{4} \\ 
\frac{3}{4} & \frac{1}{4}%
\end{array}%
\right) $ & $\frac{1}{2\sqrt{4-\sqrt{3}}}\left( 
\begin{array}{c}
\sqrt{3}-2 \\ 
3%
\end{array}%
\right) \leftrightarrow \frac{1}{4}\left( \sqrt{3}-1\right) =\lambda _{2}$
\\ \hline
$\mathbb{M}\left( 3k\right) \doteq \mathbb{M}_{3}=\left( 
\begin{array}{cc}
\frac{1}{2} & \frac{1}{2} \\ 
0 & 1%
\end{array}%
\right) ,\quad k\neq 0$ & $\left( 
\begin{array}{c}
1 \\ 
0%
\end{array}%
\right) \leftrightarrow \frac{1}{2}=\lambda _{3} $ \\ \hline
\end{tabular}%
\caption{{\small{The stochastic matrices and the decay 
modes eigenstates and eigenvalues. The PF eigenvalue is the same for 
the three cases, $\lambda _{PF}=1$.}}}
\label{Tnl2}
\end{table}
The PF eigenstates for the three matrices in Table \ref{Tnl2} are the same, 
$\frac{1}{\sqrt{2}}\left( \begin{array}{c} 1 \\ 1  \end{array} \right)$.  
\begin{table}[H]
\centering
\begin{tabular}{|c|c|}
\hline
& $T_{i} = -\left( \ln \left\vert \lambda _{i}\right\vert \right)
^{-1}$ \\ \hline\hline 
\multicolumn{1}{|l|}{$\mathbb{M}_1 ^{n}=\mathbb{C}_{1,1}+\mathbb{C}_{2,1}e^{-%
n/{T_{1}}}$} & \multicolumn{1}{c|}{$2.62 $} \\ \hline 
\multicolumn{1}{|l|}{$\mathbb{M}_2 ^{n}=\mathbb{C}_{1,2}+\mathbb{C}%
_{2,2}e^{i\left( \pi +{i}/{T_{2}}\right) n }$} & \multicolumn{1}{c|}{$%
0.59$} \\ \hline 
\multicolumn{1}{|l|}{$\mathbb{M}_3 ^{n}=\mathbb{C}_{1,3}+\mathbb{C}_{2,3}e^{-%
{n}/{T_{3}}}$} & \multicolumn{1}{c|}{$1.44$} \\ \hline
\end{tabular}%
\caption{{\small {Expansion in terms of the spectral modes and
decay times.}}}
\label{Tnl3}
\end{table}
The $\mathbb{C}_{l,i}$ matrices are given in Table \ref{Tnl4}. 
\begin{table}[H]
\centering
\begin{tabular}{|c||c|c|}
\hline
$i$ & $\mathbb{C}_{1,i}$ & $\mathbb{C}_{2,i}$ \\ \hline\hline
$1$ & $\frac{1}{\sqrt{3}+5}\left( 
\begin{array}{cc}
3 & \sqrt{3}+2 \\ 
3 & \sqrt{3}+2%
\end{array}%
\right) $ & $\frac{1}{\sqrt{3}+5}\left( 
\begin{array}{cc}
\sqrt{3}+2 & -\left( \sqrt{3}+2\right) \\ 
-3 & 3%
\end{array}%
\right) $ \\ \hline
$2$ & $\frac{1}{5-\sqrt{3}}\left( 
\begin{array}{cc}
3 & 2-\sqrt{3} \\ 
3 & 2-\sqrt{3}%
\end{array}%
\right) $ & $\frac{1}{5-\sqrt{3}}\left( 
\begin{array}{cc}
2-\sqrt{3} & -\left( 2-\sqrt{3}\right) \\ 
-3 & 3%
\end{array}%
\right) $ \\ \hline
$3$ & $\left( 
\begin{array}{cc}
0 & 1 \\ 
0 & 1%
\end{array}%
\right) $ & $\left( 
\begin{array}{rr}
1 & -1 \\ 
0 & 0%
\end{array}%
\right) $ \\ \hline
\end{tabular}%
\caption{{\small {The $\mathbb{C}_{1,i}$ matrices are stochastic, whereas 
the $\mathbb{C}_{2,i}$ matrices are not, and the sum of the entries in 
each row is null.}}}
\label{Tnl4}
\end{table}

We now assume that initially the number of vehicles in the
two cities, $A$ and $B$, are $a_0$ and $b_0$, represented by the
vector $\left( \begin{array}{cc} a_0 & b_0 \end{array} \right)$. For 
predicting the number of vehicles at three different moments of a day 
we set the rules in Table \ref{Tnl55}  
\begin{table}[ht]
\centering
\begin{tabular}{|l||c|c|}
\hline
$i$ & $r-th$ DAY evolution, $r=1,2,3,...$ & $n$ \\ \hline\hline
$1$ & $\left( 
\begin{array}{cc}
a_{r}^{\left( 1\right) } & b_{r}^{\left( 1\right) }%
\end{array}%
\right) =\left( 
\begin{array}{cc}
a_{0} & b_{0}%
\end{array}%
\right) \mathbb{M}_{1}\left( \mathbb{M}_{2}\mathbb{M}_{3}\mathbb{M}%
_{1}\right) ^{r-1}$ & $3\left( r-1 \right)+1$ \\ \hline
$2$ & $\left( 
\begin{array}{cc}
a_{r}^{\left( 2\right) } & b_{r}^{\left( 2\right) }%
\end{array}%
\right) =\left( 
\begin{array}{cc}
a_{0} & b_{0}%
\end{array}%
\right) \mathbb{M}_{1}\mathbb{M}_{2}\left( \mathbb{M}_{3}\mathbb{M}_{1}%
\mathbb{M}_{2}\right) ^{r-1}$ & $3\left( r-1 \right)+2$ \\ \hline
$3$ & $\left( 
\begin{array}{cc}
a_{r}^{\left( 3\right) } & b_{r}^{\left( 3\right) }%
\end{array}%
\right) =\left( 
\begin{array}{cc}
a_{0} & b_{0}%
\end{array}%
\right) \left( \mathbb{M}_{1}\mathbb{M}_{2}\mathbb{M}_{3}\right) ^{r}$ & $3r$
\\ \hline
\end{tabular}%
\caption{\small {Evolution of the number of vehicles, $a_{r}^{(i)}$
and $b_r^{(i)}$ at the $r$-th day, in the three stationary states $i$ associated
with sequential observations: $1$ (morning), $2$ (noon) and $3$ (evening). The
integer $n$ stands for the sequence of moments for $r$ days.}}
\label{Tnl55}
\end{table}
\noindent and by moment we mean some, arbitrarily chosen, interval of time for 
counting the number of vehicles passing by an established mark -- an intersecting 
artery -- or the traffic flux, (a sampling method) at three different hours of 
a day, for instance, morning, noon and evening, to be compared with the 
predictions of the model.

\begin{figure}[ht]
\centering
\includegraphics[height=2.4in, width=3.2in]{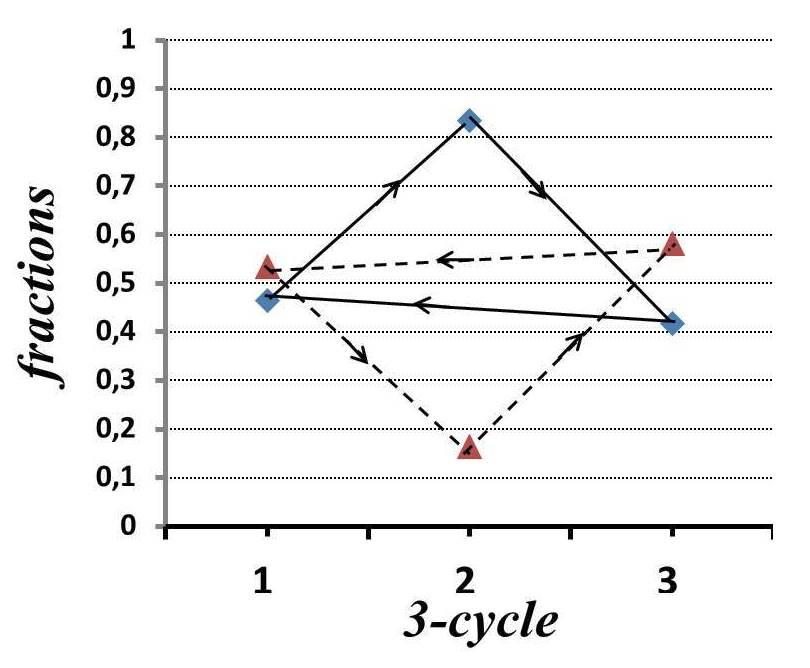}
\caption{\small {Three stationary states for each city. The flow of
vehicles changes cyclically at each step: $1$ (morning), $2$ (noon) and $3$
(evening). The arrows indicate the directions of the cyclic changes:
clockwise (solid lines) for city $A$ and counterclockwise (dashed lines) for 
$B$. The symmetry is due to the choice $a_1=a_2=1/2$ and $b_1=b_2=-1/2$ 
for the parameters of the model.}}
\label{cyclic}
\end{figure}

The properties of stochastic matrices permit to calculate their products,  
presented in Table \ref{Tnl55}, as 
\begin{subequations}
\label{matprod}
\begin{eqnarray}
\mathbb{M}_{1}\left( \mathbb{M}_{2}\mathbb{M}_{3}\mathbb{M}_{1}\right)
^{r-1} &=& \mathbb{H}_{1,1}+\mathbb{H}_{2,1}\left( -1\right) ^{r-1}e^{-\frac{%
r-1}{\tau _{1}}}\ , \\
\mathbb{M}_{1}\mathbb{M}_{2}\left( \mathbb{M}_{3}\mathbb{M}_{1}\mathbb{M}%
_{2}\right) ^{r-1} &=& \mathbb{H}_{1,2}+\mathbb{H}_{2,2}\left( -1\right)
^{r-1}e^{-\frac{r-1}{\tau _{1}}}\ , \\
\left( \mathbb{M}_{1}\mathbb{M}_{2}\mathbb{M}_{3}\right) ^{r} &=& \mathbb{H}%
_{1,3}+\mathbb{H}_{2,3}\left( -1\right) ^{r}e^{-\frac{r}{\tau _{1}}}\ , 
\end{eqnarray}
\end{subequations}
with $r=1,2,...$, and since the eigenvalues of the three matrices (\ref{matprod}) 
coincide, $\lambda_1 = -(1/16)$, the decay time is $\tau_1 =
-\left(\ln\left(1/16\right)\right)^{-1} \approx 0.36$. The $\mathbb{H}$
matrices are 
\begin{subequations}
\begin{eqnarray}
\mathbb{H}_{1,1} &=&\frac{1}{34}\left( 
\begin{array}{cc}
21-3\sqrt{3} & 13 +3\sqrt{3} \\ 
21-3\sqrt{3} & 13+3\sqrt{3}%
\end{array}%
\right) \approx \left( 
\begin{array}{cc}
0.46\, & 0.54\, \\ 
0.46\, & 0.54\,%
\end{array}%
\right) \\
\mathbb{H}_{2,1} &=&\frac{1}{17}\left( 
\begin{array}{rc}
-\left( \frac{11}{4}\sqrt{3}+2\right) & \frac{11}{4}\sqrt{3}+2 \\ 
\frac{3}{2}\left( \sqrt{3}+\frac{3}{2}\right) & -\frac{3}{2}\left( \sqrt{3}+%
\frac{3}{2}\right)%
\end{array}%
\right) \approx \left( 
\begin{array}{rr}
-0.40\, & 0.40\, \\ 
0.29\, & -0.29\,%
\end{array}%
\right)
\end{eqnarray}
\end{subequations}
\begin{subequations}
\begin{eqnarray}
\mathbb{H}_{1,2} &=&\frac{1}{17}\left( 
\begin{array}{cc}
9+3\sqrt{3} & 8-3\sqrt{3} \\ 
9+3\sqrt{3} & 8-3\sqrt{3}%
\end{array}%
\right) \approx \left( 
\begin{array}{cc}
0.84\, & 0.16\, \\ 
0.84\, & 0.16\,%
\end{array}%
\right) \\
\mathbb{H}_{2,2} &=&\frac{1}{272}\left( 
\begin{array}{rr}
-25+3\sqrt{3} & 25-3\sqrt{3} \\ 
3\left( \sqrt{3}+3\right) & -3\left( \sqrt{3}+3\right)%
\end{array}%
\right) \approx \left( 
\begin{array}{rr}
-0.07 & 0.07 \\ 
0.05 & -0.05%
\end{array}%
\right)
\end{eqnarray}
\end{subequations}
\begin{subequations}
\begin{eqnarray}
\mathbb{H}_{1,3} &=&\frac{1}{34}\left( 
\begin{array}{rr}
3\sqrt{3}+9 & 25-3\sqrt{3} \\ 
3\sqrt{3}+9 & 25-3\sqrt{3}%
\end{array}%
\right) \approx \left( 
\begin{array}{rr}
0.42 & 0.58\, \\ 
0.42 & 0.58\,%
\end{array}%
\right) \\
\mathbb{H}_{2,3} &=&\frac{1}{34}\left( 
\begin{array}{rr}
25-3\sqrt{3} & -25+3\sqrt{3} \\ 
-\left( 3\sqrt{3}+9\right) & 3\sqrt{3}+9%
\end{array}%
\right) \approx \left( 
\begin{array}{rr}
0.58 & -0.58 \\ 
-0.42 & 0.42%
\end{array}%
\right)\ .
\end{eqnarray}

Thus, asymptotically the average number of vehicles associated with each city
(whose distribution will repeat daily) at the three moments of one day are 
\end{subequations}
\begin{eqnarray}
\left( 
\begin{array}{cc}
a_{\infty }^{\left( 1\right) } & b_{\infty }^{\left( 1\right) }%
\end{array}%
\right) &=&\lim_{r\rightarrow \infty }\left( 
\begin{array}{cc}
a_{0} & b_{0}%
\end{array}%
\right) \mathbb{M}_{1}\left( \mathbb{M}_{2}\mathbb{M}_{3}\mathbb{M}%
_{1}\right) ^{r-1}=\left( 
\begin{array}{cc}
a_{0} & b_{0}%
\end{array}%
\right) \mathbb{H}_{1,1}  \notag \\
&=& \left( a_{0}+b_{0}\right)\ \left( 
\begin{array}{rr}
\frac{3}{34}\left( 7-\sqrt{3}\right) & \frac{1}{34}\left( 3\sqrt{3}+13\right)%
\end{array}%
\right)  \notag \\
&\approx & \left( a_{0}+\,b_{0}\right)\ \left( 
\begin{array}{cc}
0.46 & 0.54%
\end{array}%
\right) \ , \label{Rnl3}
\end{eqnarray}
\begin{eqnarray}
\left( 
\begin{array}{cc}
a_{\infty }^{\left( 2\right) } & b_{\infty }^{\left( 2\right) }%
\end{array}%
\right) &=&\lim_{k\rightarrow \infty }\left( 
\begin{array}{cc}
a_{0} & b_{0}%
\end{array}%
\right) \mathbb{M}_{1}\mathbb{M}_{2}\left( \mathbb{M}_{3}\mathbb{M}_{1}%
\mathbb{M}_{2}\right) ^{r}=\left( 
\begin{array}{cc}
a_{0} & b_{0}%
\end{array}%
\right) \mathbb{H}_{1,2}  \notag \\
&=& \left( a_{0}+b_{0}\right) \left( 
\begin{array}{rr}
\frac{3}{17}\left( \sqrt{3}+3\right) & \frac{1}{17}\left( 8-3\sqrt{3}\right)%
\end{array}%
\right)  \notag \\
&\approx & \left( a_{0}+\,b_{0}\right) \left( 
\begin{array}{cc}
0.84 & 0.16%
\end{array}%
\right) \ , \label{Rnl4}
\end{eqnarray}
\begin{eqnarray}
\left( 
\begin{array}{cc}
a_{\infty }^{\left( 3\right) } & b_{\infty }^{\left( 3\right) }%
\end{array}%
\right) &=&\lim_{k\rightarrow \infty }\left( 
\begin{array}{cc}
a_{0} & b_{0}%
\end{array}%
\right) \left( \mathbb{M}_{1}\mathbb{M}_{2}\mathbb{M}_{3}\right) ^{r}=\left( 
\begin{array}{cc}
a_{0} & b_{0}%
\end{array}%
\right) \mathbb{H}_{1,3}  \notag \\
&=& \left( a_{0}+b_{0}\right) \left( 
\begin{array}{cc}
\frac{3}{34}\left( \sqrt{3}+3\right) & \frac{1}{34}\left( 25-3\sqrt{3}\right)%
\end{array}%
\right)  \notag \\
&\approx & \left( a_{0}+\,b_{0}\right) \left( 
\begin{array}{cc}
0.42 & 0.58%
\end{array}%
\right)\ . \label{Rnl5}
\end{eqnarray}
In these last three equations (third lines) the numbers in each pair $(0.46,0,54)$, 
$(0.84,0.16)$ and $(0.42,0.58)$ relate to the fractions, of all vehicles, $a_0+b_0$, 
associated with each city at different moments. For instance, the first pair 
(morning monitoring) means that $46\%$ of all the vehicles are either in city 
$A$ or driving toward city $B$, while $54\%$ are in $B$ or driving toward $A$. 
The same holds for the two other pairs, although being for noon and evening 
respectively. These fractions change periodically and continuously, as schematized 
in Fig. \ref{cyclic}, where the arrows indicate the direction of the flow of vehicles. 
The results do not depend on the initial values $a_0$ and $b_0$ individually,
but only on their sum, so some information is lost. The percentages depend only 
on the chosen parameters of the stochastic matrix. 
The flow of vehicles change cyclically, ruled by the stationary matrices 
$\mathbb{H}_{1,i}$ and $\mathbb{H}_{2,i}$ as displayed in Fig. \ref%
{cyclicH}. 
\begin{figure}[H]
\centering
\includegraphics[height=1.2in, width=1.6in]{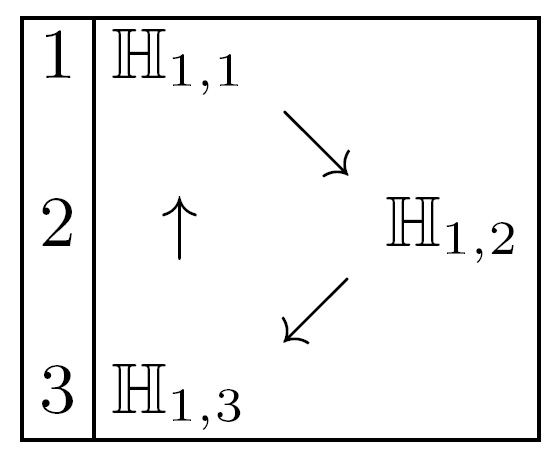}
\caption{\small {The stationary traffic flow changes cyclically
according to the asymptotic matrices $\mathbb{H}_{1,i}$, and depending on the
moments of the day: $1$ (morning), $2$ (noon) and $3$ (evening).}}
\label{cyclicH}
\end{figure}
%
%
The global mean entropy also presents asymptotic regular cyclical changes 
as shown in Fig. \ref{NLentropy}, which is in line with Figs. \ref{cyclic} 
and \ref{cyclicH}.
\begin{figure}[htbp]
\centering
\includegraphics[height=2.0in, width=3.0 in]{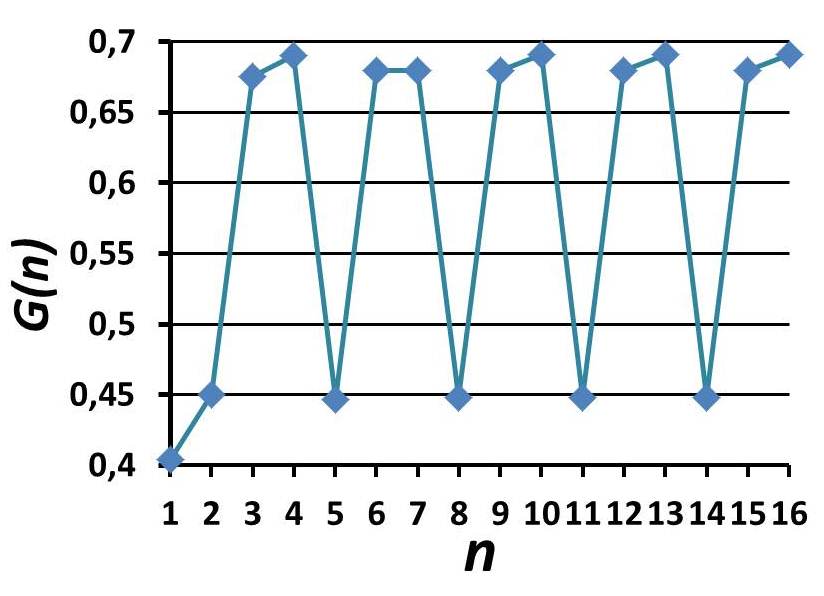}
\caption{\small {The global mean entropy for the nonlinear model as function of $n$. 
The solid line only links the points and the 3-cycle signature is the pattern of the 
periodic repetition.}}
\label{NLentropy}
\end{figure}

As so, this particular nonlinear model leads to three stationary matrices
(three fixed points in the language of dynamical systems) characterizing a
3-cycle continuous change of the vehicular traffic, that we assumed as being 
three moments of observation every day. This approach can be extended to $L$ 
stationary states at $L$ different moments. The model is scalable by increasing 
the number of sites and arteries that connect them, and the obtainment of 
numerical results depends only on computational capabilities. To construct 
a model that is more kin to the real traffic within a network it is advisable 
to insert the dynamics of input and output of vehicles (as considered in the 
previous model), thus turning it into a hybrid model.
%
%
\section{Summary and conclusions}
%
We presented three network models to picture the vehicular traffic  
between sites that could be cities, parking lots or car-rental agencies, 
and arteries (highways, roads) that connect the sites for the circulation 
of vehicles. We opted to use the mathematical formalism based on stochastic 
matrices to simulate the evolution in time of the distribution of vehicles. 
By doing a convenient decomposition of the dynamically 
evolved stochastic matrix in several modes we have separated the stationary 
matrix from the transient ones, and for these we have defined characteristic 
decay times. The first model considered a network without sources and sinks 
for the circulation of vehicles (a closed system) and using a numerical 
example we verified that after very few steps of evolution the vehicular 
distribution attains the stationary state. The second model consists of the 
same previous network accreted with a source and a sink, i.e., two additional 
sites and arteries. One extra site contains a certain number of new vehicles 
that are inserted daily into the network at a given rate, and the other extra 
site is a depository of old and crashed vehicles removed from 
the network, at an also established rate. Depending on numerical values chosen 
as entries of the SM's, the main differences that result from the former model 
are: (1) very long decay times to attain the stationary state, where all the 
vehicles will go eventually to the depository site; (2) the evolution of the 
global mean entropy begins by increasing, then it attains a maximum value 
that decreases steeply, reaching zero soon after. We believe that this model 
could help urban planners to establish what could be the ideal injection of 
vehicles within a network in order to avoid an excessive density that could 
saturate the free circulation capacity within the arteries.

The third model, for two sites and two arteries connecting them, is nonlinear. 
Comparing the properties with the linear models, the differences are quite 
noteworthy: instead of a single stationary state, now multi-stationary states 
are possible and each mode has its own relaxation time although the PF eigenvalue 
($\lambda_{PF} = 1$) is always present, even for the $n$-step evolution. We 
chose specific sinusoidal functions of $n$ in the entries of the SM, and 
asymptotically one gets $L$ stationary states. Therefore, the model admits the 
possibility to describe a particular situation for the traffic flow: $L$ different 
number of vehicles at each site, changing cyclically in one day. We consider that 
this model is more realistic to describe the possible variations of the traffic 
flow at different moments of the day. 

Finally, we recall that all three models are scalable, i.e., depending on 
the computational resources, the network can be extended to a large number 
of vertices and edges. A following paper (Part II) makes use of the formalism 
and models discussed here to analyze of a real situation, the urban traffic 
within a sector of Tigre, a city located in the province of Buenos Aires, 
Argentina. The recorded traffic by video cameras was made available by the 
traffic controllers. 
%
\begin{acknowledgements}
SSM thanks the CNPq, a Federal Brazilian Agency, for financial support.
\end{acknowledgements}

%

%
\end{document}